\documentclass{article}
\usepackage{ijcai19}

\usepackage{times}
\usepackage{soul}
\usepackage{url}
\usepackage[hidelinks]{hyperref}
\usepackage[utf8]{inputenc}
\usepackage[small]{caption}
\usepackage{microtype}
\usepackage{graphicx}
\usepackage{algcompatible}
\usepackage{subfigure}
\usepackage{amsfonts}
\usepackage{amssymb}
\usepackage{amsmath}
\usepackage{multirow}
\usepackage{comment}
\usepackage{balance}
\usepackage{booktabs}
\usepackage[ruled]{algorithm}
\title{Pipeline Parallelism for Inference on Heterogeneous Edge Computing}

\author{
Yang Hu$^{1,2}$\and
Connor Imes$^2$\and
Xuanang Zhao$^{3}$\and
Souvik Kundu$^1$\and
Peter A. Beerel$^1$\and
Stephen P. Crago$^2$\and
John Paul N. Walters$^2$\\
\affiliations
$^1$Viterbi School of Engineering,
University of Southern California\\
$^2$Information Sciences Institute,\textbf{}
University of Southern California\\
$^3$Google\\
\emails
yhu21003@usc.edu,
cimes@isi.edu,
sd.invol@gmail.com,
\{souvikku, pabeerel\}@usc.edu,
\{crago, jwalters\}@isi.edu
}

\begin{document}

\maketitle

\begin{abstract}
Deep neural networks with large model sizes achieve state-of-the-art results for tasks in computer vision (CV) and natural language processing (NLP). However, these large-scale models are too compute- or memory-intensive for resource-constrained edge devices. Prior works on parallel and distributed execution primarily focus on training---rather than inference---using homogeneous accelerators in data centers.
% uses model compression to shrink model sizes and accelerate inference, which requires iterative retraining and reduces accuracy, or
We propose EdgePipe, a distributed framework for edge systems that uses pipeline parallelism to both speed up inference and enable running larger (and more accurate) models that otherwise cannot fit on single edge devices. EdgePipe achieves these results by using an optimal partition strategy that considers heterogeneity in compute, memory, and network bandwidth. Our empirical evaluation demonstrates that EdgePipe achieves $10.59\times$ and $11.88\times$ speedup using 16 edge devices for the ViT-Large and ViT-Huge models, respectively, with no accuracy loss. Similarly, EdgePipe improves ViT-Huge throughput by $3.93\times$ over a 4-node baseline using 16 edge devices, which independently cannot fit the model in memory.  Finally, we show up to $4.16\times$ throughput improvement over the state-of-the-art PipeDream when using a heterogeneous set of devices.

\end{abstract}

\section{Introduction}

\begin{figure}[t]
    \centering
    \includegraphics[width=0.8\linewidth]{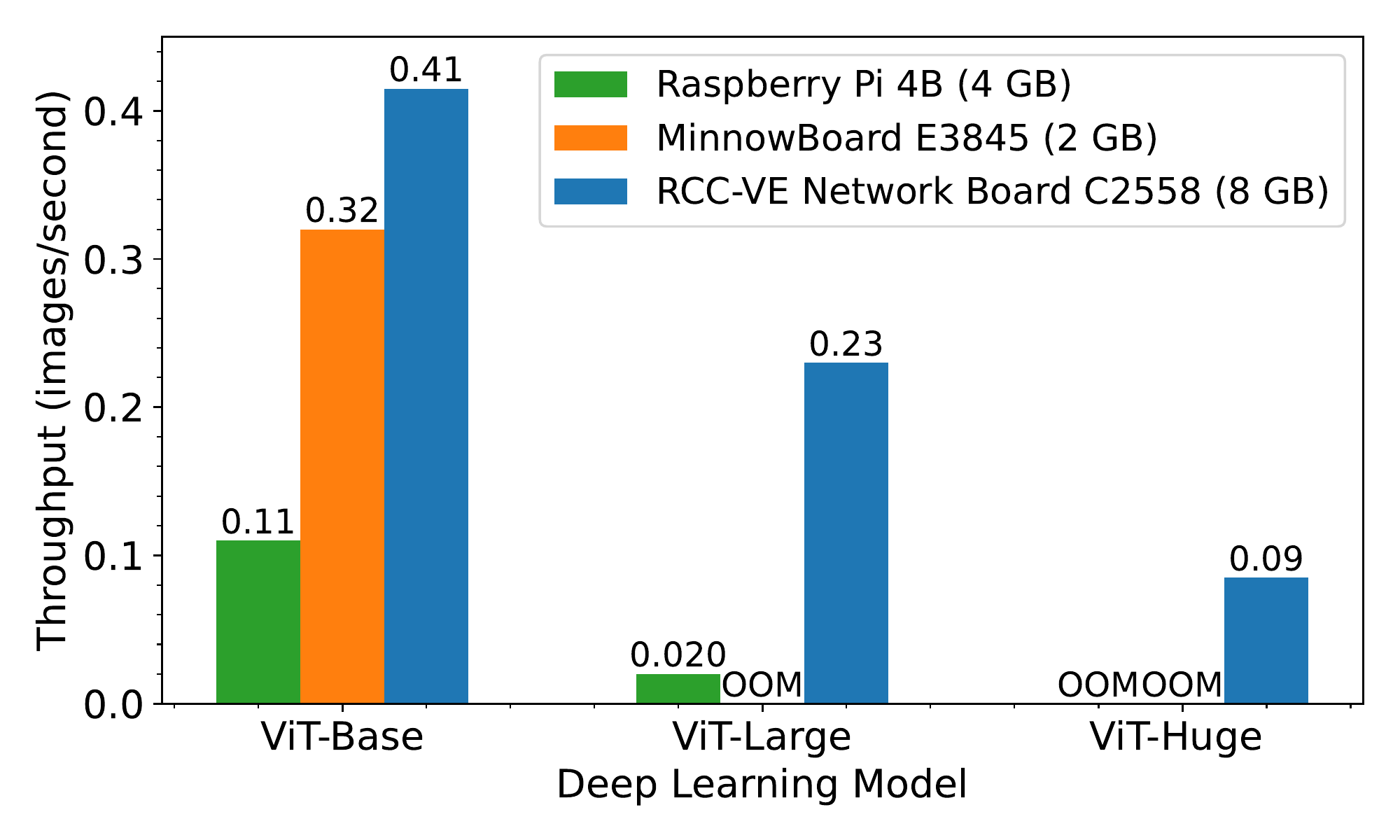}
    \vspace{-4mm}
    \caption{ViT model performance on three edge devices. \texttt{OOM} indicates the model does not fit in memory.}
    \label{fig:intro_model}
    \vspace{-4mm}
\end{figure}

In recent years, deep neural network (DNN) model sizes have increased exponentially to provide better accuracy \cite{krizhevsky2012imagenet,redmon2017yolo9000,tao2018image}. In particular, large transformer-based models have achieved state-of-the-art accuracy in various computer vision (CV) \cite{dosovitskiy2020image,yuan2021tokens} and natural language processing (NLP) tasks \cite{vaswani2017attention,dosovitskiy2020image,carion2020end}. However, these large models cause significant challenges for training and inference in all environments, especially at the edge, which consists of resource-constrained devices in close proximity to a data source \cite{edgecomputing}.
%Thus, despite their impressive performance, their deployment remains limited to devices with large compute budgets, excluding many edge devices.   
For example, the base vision transformer model (ViT-Base) \cite{dosovitskiy2020image} has 86.6M  parameters and requires about 110B FLOPs to perform inference on one image \cite{narayanan2021efficient}, resulting in limited throughput on a MinnowBoard edge device, as shown in Figure~\ref{fig:intro_model}. The ViT-Large and ViT-Huge models have considerably more parameters, which makes their deployment on resource-constrained edge devices more difficult, e.g., they do not even fit in memory on the MinnowBoard.

Various methods have been proposed to address large model inference challenges on edge devices, including model compression \cite{hinton2015distilling,han2015deep,yao2021hawq,kundu2021attentionlite}, adaptive inference \cite{tambe2020edgebert}, and neural architecture search \cite{hat}. These approaches %provide feasible solutions for deploying deep learning models on edge devices by 
reduce the number of required computation operations, but at the cost of reduced accuracy. Moreover, most of these approaches are limited to a single device, and do not take advantage of idle devices that may be available to assist and improve performance in distributed settings.

Pipeline parallelism partitions models into multiple stages, which can accelerate processing without accuracy loss by utilizing additional resources. %to assist in inference. %More precisely, this method partitions a large model into multiple parts and utilizes the extra computational resources from other edge or cloud devices to execute distributed  inference.
Pipelining is complementary to compression methods, providing additional opportunities to mitigate model complexity.
Research on pipeline parallelism has focused on data center scenarios with high interconnect bandwidth and homogeneous accelerators like graphics processing units (GPUs) and tensor processing units (TPUs) \cite{pipemare,huang2019gpipe,li2021terapipe,he2021pipetransformer}. In contrast, edge environments are more resource-constrained, with heterogeneous communication and computation characteristics.

Several other frameworks consider pipeline parallelism for limited heterogeneity in data centers, e.g., heterogeneous communication topologies with homogeneous GPUs \cite{pipedream,pipedream_2bw}, or heterogeneous GPU clusters with homogeneous networks \cite{hetpipe}. Torchpipe \cite{torchgpipe} provides an automatic balancing strategy for large models, but only for the single-node scenario and does not claim optimality. Finding an optimal partition strategy under fully heterogeneous conditions (heterogeneous devices and network), which is critical to edge scenarios, remains a largely open problem.

We address these challenges with EdgePipe, a distributed inference framework that exploits pipeline parallelism to improve inference performance on heterogeneous edge devices and networks. In particular, this paper makes the following contributions:
\begin{itemize}
    \item A distributed pipeline parallelism framework to accelerate large-scale model inference on heterogeneous edge computing without accuracy loss.
    
    \item A dynamic programming (DP) algorithm to determine the optimal partition mapping of pipeline parallelism to heterogeneous devices and communication channels. 

    \item A detailed experimental evaluation on a real edge testbed, demonstrating throughput performance improvements up to $11.88\times$ speedup in a 16-device homogeneous cluster and $4.16\times$ speedup over the state-of-the-art PipeDream in a heterogeneous cluster\footnote{To support reproducibility, we will open-source our evaluation code upon acceptance of the paper.}. 
    
\end{itemize}
\section{Background and Motivation}
\label{background_and_motivation}
%%%%%%%%%%%%%%%%%%%%%%%%%%%%%%%%%% Background Start %%%%%%%%%%%%%%%%%%%%%%%%%%%%%%%%%%%%
%%%%%%%%%%%%%%%%%%%%%%%%%%%%%%%%%%%%%%%%%%%%%%%%%%%%%%%%%%%%%%%%%%%%%%%%%%%%%%%%%%%%%%%%
\subsection{Background}

\textbf{Pipeline Parallelism}. Pipeline parallelism partitions a neural network model into multiple stages, where each stage consists of a consecutive set of layers in the original model \cite{pipemare,pipedream}. Each stage in the pipeline is assigned to a worker to achieve the parallel execution of model training or inference. The input minibatch is split into multiple chunks of equal size, which are called microbatches \cite{huang2019gpipe}. The microbatch size affects the pipeline performance, with the optimal size depending on multiple factors including the characteristics of the model and the number of pipeline stages \cite{narayanan2021efficient}. A worker in a system that has pipeline parallelism need only send its output data to a single worker, which avoids the collective communication to synchronize results with all workers. Pipelining can also overlap computation and communication to improve the performance \cite{pipedream}.

\textbf{Transformer-based Models}. The transformer model was proposed to improve the effectiveness in learning dependencies between distant positions for sequence modeling tasks \cite{vaswani2017attention}. A transformer encoder includes multiple transformer layers with identical structures. Every transformer layer is composed of a multi-head self-attention layer, a multi-layer perceptron (MLP), two layer normalization operations, and residual connections. The multi-head self-attention layer calculates the attention score of the input sequence $A = (a_1, ..., a_n)$ through dot product operations and generates the output representation $B=(b_1, .., b_n)$ with the same dimension. Outputs $b_i$ and $b_j$ (where $i \neq j$) are operationally independent, which provides the possibility of parallel execution for both training and inference \cite{park2020optimus}. Attention-based models have recently been extended to replace conventional convolutional neural networks (CNNs) \cite{dosovitskiy2020image,kundu2021attentionlite} in performing complex CV tasks. In particular, the ViT models enjoy superior representation ability \cite{d2021convit}, and also suffer less from positional invariance issues which are prevalent in conventional CNNs \cite{su2019pixel}.

\begin{figure*}[!th]
    \centering
    \includegraphics[width=0.97\linewidth]{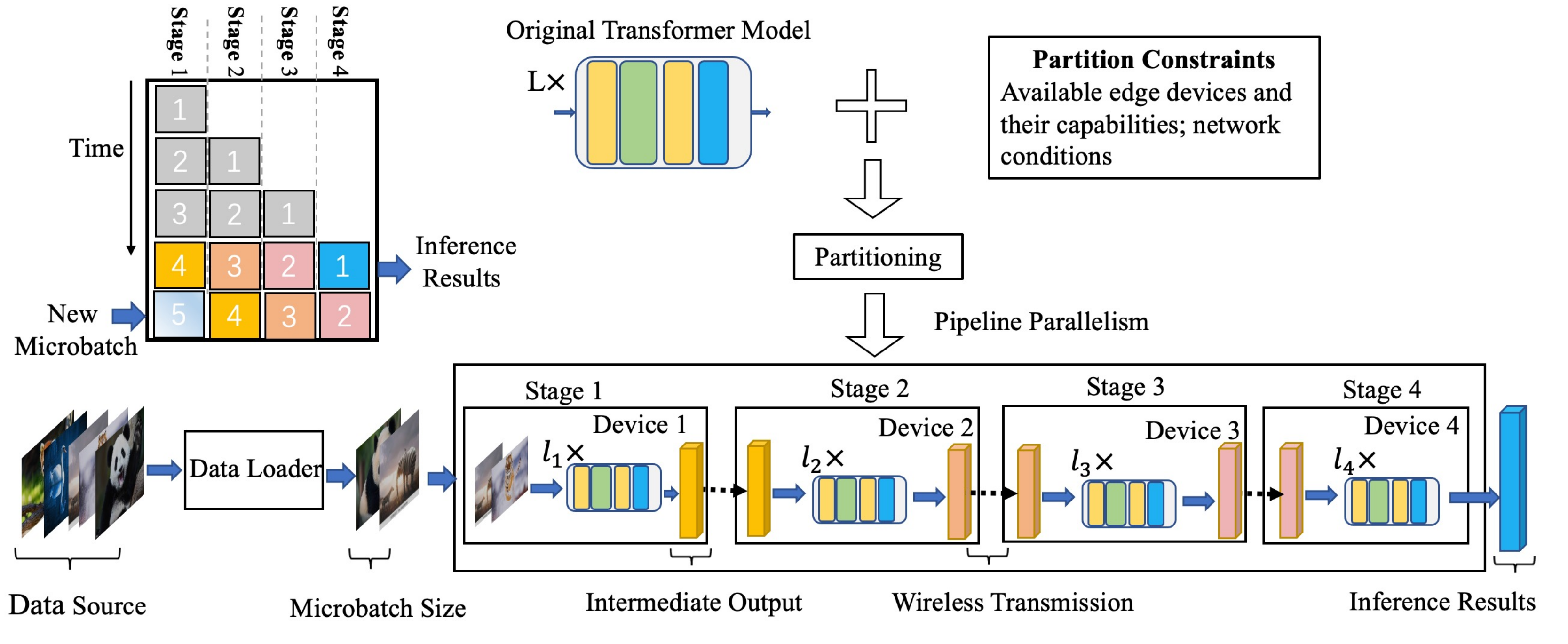}
    \vspace{-4mm}
    \caption{EdgePipe system design overview. EdgePipe automatically partitions transformer models between edge devices, subject to device, network, and model constraints. Intermediate results then are transmitted between pipeline stages. Figure best seen in color.}
    \label{fig:system_overview}
\end{figure*}

\subsection{Motivation}
Large-scale model inference is a challenging task for resource-limited devices. Classical model compression techniques, including pruning \cite{zhang2018systematic,kundu2021dnr}, quantization \cite{yao2021hawq}, low-rank approximation \cite{chin2020towards}, and knowledge distillation \cite{hinton2015distilling} can shrink neural network model sizes to potentially accelerate deep neural networks.  However, these techniques often require iterative retraining and a full-precision pre-trained model to avoid significant accuracy-drop. More importantly, these methods generally focus on a single compute node. Distributed edge computing scenarios, in contrast, often include a large number of resource-limited devices, e.g., in vehicular edge computing (VEC) for internet of vehicles \cite{vec}, wireless-connected AI-enabled sensors \cite{wireless_sensor_network}, and smart home systems \cite{smart_speaks,smart_home_survey}.

Pipeline parallelism has proven to be effective for distributed training on accelerators in data centers where devices are relatively homogeneous and the network bandwidth is generally high \cite{huang2019gpipe,pipemare,he2021pipetransformer}. However, edge computing has unique characteristics:
\begin{itemize}
    \item \textit{Small Memory Capacity}. Compared with data-center servers, edge devices usually have smaller memory capacities, ranging from tens of MB to several GB. 
    %Partitioned model shards must conform to the memory limitations of edge devices.  
    \item \textit{Heterogeneous Devices}. Edge devices have diverse computational performance and memory capacities.
    \item \textit{Limited Bandwidth}. Unlike communication within data centers, edge computing systems often rely on wireless communication with limited bandwidth. 
    \item \textit{Heterogeneous Network Link Capacities}. The bandwidth between different pairs of devices depends on physical distance and channel interference, and could range from tens of Kbps to hundreds of Mbps.
\end{itemize}

% Due to this heterogeneity,
Model partition methods for homogeneous clusters will therefore perform poorly in heterogeneous edge environments.
A new pipeline parallelism framework is needed to overcome these challenges. % in heterogeneous edge environments.
In the next section, we introduce a framework for distributed edge clusters that uses heterogeneity-aware pipeline parallelism to improve inference performance and enable running larger---and more accurate---models.

%%%%%%%%%%%%%%%%%%%%%%%%%%%%%%%%%% Background End %%%%%%%%%%%%%%%%%%%%%%%%%%%%%%%%%%%%%%
%%%%%%%%%%%%%%%%%%%%%%%%%%%%%%%%%%%%%%%%%%%%%%%%%%%%%%%%%%%%%%%%%%%%%%%%%%%%%%%%%%%%%%%%

\section{Parallelism for Edge Devices}
\label{parallelism}
%%%%%%%%%%%%%%%%%%%%%%%%%%%%%%%%%% Parallelism Start %%%%%%%%%%%%%%%%%%%%%%%%%%%%%%%%%%%
%%%%%%%%%%%%%%%%%%%%%%%%%%%%%%%%%%%%%%%%%%%%%%%%%%%%%%%%%%%%%%%%%%%%%%%%%%%%%%%%%%%%%%%%

In this section, we present our system design and discuss the problem of partitioning a transformer model for a fully heterogeneous cluster. We introduce the DP-based optimal partition algorithm in Section~\ref{sec:optimization}.  

\subsection{EdgePipe System Design}

Figure~\ref{fig:system_overview} presents the EdgePipe system design, which consists of three major components: the partitioning algorithm for heterogeneous clusters, the data loader, and the runtime framework that implements pipeline parallelism. First, the configuration of the original transformer model and partition constraints are sent to the partition algorithm to generate an optimal partition method. The partition constraints include available edge devices,  computation and memory capabilities of these devices, and the bandwidth between devices. The specific partition problem will be introduced in Section \ref{sec:algorithm}. To fit into every edge device and improve the throughput, input data should be split into small chunks, called a microbatch.  At runtime, selected devices are only responsible for the inference of one part of the original model. After finishing the inference of one microbatch, the edge device transmits intermediate outputs to the device in the next pipeline stage. The device in the final stage produces the final result, which could be transmitted to another host or stored locally.

To construct a pipeline with $k$ stages, the transformer model should be partitioned into $k$ parts $T_1, T_2, ...,T_k$. Part $T_i$ includes $l_i$ layers, and $\sum_{i=1}^{k} l_i = L$, where $L$ is the number of layers. Part $T_i$ is assigned to the $i$-th device to construct the $i$-th pipeline stage. For one microbatch input $\mathcal{X}$, the process of inference may be denoted as $\mathcal{\hat{Y}} = T_{k}(T_{k-1}...(T_2(T_1(\mathcal{X}))))$, where $\mathcal{\hat{Y}}$ is the final output of the inference. The intermediate output of the $i$-th device is sent to the $(i+1)$-th device to continue the computation. The number of pipeline inference stages in EdgePipe is equal to the number of devices participating in inference. 

%%%%%%%%%%%%%%%%%%%%%%%%%%%%%%%%%% Parallelism End %%%%%%%%%%%%%%%%%%%%%%%%%%%%%%%%%%%%%
%%%%%%%%%%%%%%%%%%%%%%%%%%%%%%%%%%%%%%%%%%%%%%%%%%%%%%%%%%%%%%%%%%%%%%%%%%%%%%%%%%%%%%%%
\subsection{Partition Problem Formulation}
\label{sec:algorithm}

%%%%%%%%%%%%%%%%%%%%%%%%%%%%%%% Partitioning Start %%%%%%%%%%%%%%%%%%%%%%%%%%%%%%%%%%%%%
%%%%%%%%%%%%%%%%%%%%%%%%%%%%%%%%%%%%%%%%%%%%%%%%%%%%%%%%%%%%%%%%%%%%%%%%%%%%%%%%%%%%%%%%
Fully heterogeneous clusters include heterogeneity in both the devices and the communication networks. It is common in edge computing for devices have different computation and memory capabilities. In addition, the network bandwidth between devices may be different. It is therefore challenging to decide the partition method for the cluster.
% Figure~\ref{fig:pipeline_paritition} demonstrates a concrete example.

We define a transformer model $\mathbb{T}$ with $L$ layers, inter-layer transmission data size $P_j$ for the $j$-th layer, and a list of heterogeneous devices $\mathbb{D}$ ($|\mathbb{D}| = D$) with different memory, computation, and communication capabilities. In heterogeneous communication, the bandwidth between a pair of devices $u$ and $v$ may be different than the bandwidth between a different pair of devices $u'$ and $v'$: $b_{u, v} \neq b_{u', v'}$. The optimal strategy $\mathbb{R}$ partitions the model $\mathbb{T}$ into $S$ parts and allocates them to the selected devices $\mathbb{S} \subseteq \mathbb{D}, |\mathbb{S}| = S \leq D$ to achieve maximal throughput and conform to the memory limitations of the selected devices.

\subsection{Target Optimization}
\label{sec:optimization}
% To find the optimal strategy strategy, we assume the computation time for each layer on the specific device is known.
We denote $T_{comp} (l , u)$ as the execution time for the set of layers $l$ on a device $u$. $T_{comm} (u, v, P_j)$ is the time to communicate data $P_j$ between devices $u$ and $v$, which is computed by Equation \eqref{T_comm}, where $b_{u, v}$ is the bandwidth between devices $u$ and $v$.
\begin{equation}
    T_{\text{comm}} (u, v, P_j) = \frac{P_j}{b_{u, v}}
\label{T_comm}
\end{equation}

\begin{table}
  \caption{Symbol definitions}
  \label{tab:freq}
  \small
  \begin{minipage}{\columnwidth}
  \begin{center}
  \begin{tabular}{p{0.3\linewidth} | p{0.60\linewidth}}
    \toprule
    Symbol & Description\\
    \midrule
    $\mathbb{T}, L, P_j, M_j$      & A transformer model with $L$ layers. The $j$-th layer has $P_j$ parameters for transmission and requires $M_j$ runtime memory for execution. \\
    $\mathbb{D}, D, m_v$      & A list of $D$ available devices. Device $v$ has the memory capacity $m_v$. \\
    $\mathbb{S}, S$       & The list of $S$ selected devices. Every device should participate in the inference.  \\
    $\mathbb{B}, b_{u,v}$       & Bandwidth between devices. $b_{u,v}$ is the bandwidth between devices $u$ and $v$  \\
     $\mathbb{R}$      &  The optimal mapping strategy. \\
    $T_{\text{comp}}(l, u)$     &  Computation time with the set of layers $l$ on the device $u$. \\ 
    $T_{\text{comm}}(u, v, P_j)$  & Communication time for transferring $P_j$ data from devices $u$ to $v$.  \\ 
    $T_{\text{period}}(l, u, v, P_j)$  & The maximum latency of executing pipeline stage on device $u$ with transferring $P_j$ data to device $v$.  \\ 
    $T_{\text{opt}}$  & The optimal time for the pipeline stage under given conditions. \\\hline
  \bottomrule
\end{tabular}
\end{center}
\end{minipage}
\end{table}
We assume the pipeline system supports asynchronous communication, and the computation time and communication time are perfectly overlapped. Thus, the maximum latency for the single device $u$  can be calculated as:
\begin{equation}
    T_{\text{period}} (l, u, v, P_j) = \max 
    \left\{
             \begin{array}{lr}
             T_{\text{comp}} (l, u)\\
 
             T_{\text{comm}} (u, v, P_j) &  
             \end{array}
\right. \label{T_Period}
\end{equation}
For the selected devices, achieving the maximal throughput is equivalent to minimizing the execution time $T_{\text{opt}}$, which is determined by the slowest stage under the given strategy and is equal to the largest $T_{\text{period}} (l, u, v, P_j)$. The problem of pipeline partitioning can itself be partitioned. The optimal solution for partitioning the whole pipeline on given set of devices can be constructed from the optimal partitioning result for the sub-problem, which could be solved by DP methods.

To tackle this partition problem for fully heterogeneous clusters, we design a three-dimensional DP algorithm recording the state of processed layers, used devices, and the device in the last pipeline stage. Let $h(i, \mathbb{S}, u)$ denote the minimum time to process the first $i$ layers with the set of used devices $\mathbb{S}$, and the device $u$ is the next device 
to be used. $h(i, \mathbb{S}, u)$ is the optimal solution of the subproblem for $i$ layers and $\mathbb{S}$ devices. The final optimal solution of this partition problem is the minimum $T_{\text{opt}} = h(L, \mathbb{S}, \varnothing)$ with $\mathbb{S} \subseteq \mathbb{D}$.

%% consist of two notations
The calculation of $h(j, \mathbb{S}\cup \{u\}, v)$ needs to use the optimal subproblem property: it is determined by the previous state $h(i, \mathbb{S}, u), 0 \leq i < j \leq L$,  or the calculation time $T_{\text{period}}(\{i \rightarrow j\}, u, v,P_j)$ from $i$-th layer to $j$-th layer on the current device $u$. We further analyze these two situations:
\begin{itemize}
   \item $h(j, \mathbb{S}\cup \{u\}, v) \leftarrow h(i, \mathbb{S}, u)$, the slowest pipeline stage for $j$ transformer layers is determined by the previous stage $h(i, \mathbb{S}, u)$. Since device $u$ implements the current stage from the $i$-th to the $j$-th layer in the current pipeline, the used devices set for the next state $h(j, \mathbb{S}\cup \{u\}, v)$ should include the device $u$, noted as $\mathbb{S} \cup \{u\}$.  Parameters $i$ and $u\in \mathbb{D} \setminus \mathbb{S}$ will be enumerated to find the optimal solution of the subproblem. 
   \item $h(j, \mathbb{S}\cup \{u\}, v) \leftarrow T_{\text{period}}(\{i\rightarrow j\}, u,v ,P_j)$, the device $u$ that constitutes the slowest stage of the current pipeline and limits the performance of the system. The $T_{\text{period}}(\{i \rightarrow j\}, u,v,P_j)$ is calculated from Equation~\ref{T_Period}. Similarly, device $u$ and first $i$ layers are enumerated to obtain the minimum value.
 \end{itemize} 
Thus, the state transition equation can be formulated as:  
\begin{equation}
\begin{split}
\underset{\mathbb{S}'=\mathbb{S}\cup \{u\}}{h(j, \mathbb{S}', v)} &= \min_{\substack{0\leq i < j \leq L \\ u,v \in \mathbb{D} \setminus \mathbb{S}}}  \max 
\left\{
             \begin{array}{lr}
             h(i, \mathbb{S}, u) \\
             T_{\text{period}} (\{i \rightarrow j\}, u,v, P_j)
             \end{array}
 \right. \\
&= \min_{\substack{0\leq i < j \leq L \\ u,v \in \mathbb{D} \setminus \mathbb{S}}}  \max 
\left\{
             \begin{array}{lr}
             h(i, \mathbb{S}, u)\\
             T_{\text{comm}}(u ,  v, P_j) &  \\
             T_{\text{comp}}(\{i \rightarrow j\}, u)  &  
             \end{array}                
\right. \label{transistion_equation}
\end{split}
\end{equation}
where the first term inside the max is the minimum time for the first $i$ layers with the set of devices $\mathbb{S}$ and the next used devices $u$; the second term is communication time of transferring $P_j$ data from device $u$ to device $v$; the third term is the computation time for the last $j-i$ layers on the device $u$. For initialization, $h(0, \emptyset, \varnothing)$ is set to $0$.

Equation~\ref{transistion_equation} calculates the optimal pipeline execution time. However, we need to obtain the selected devices and their order in the pipeline for the optimal strategy. Algorithm~\ref{strategy} describes the memoization technique pseudo-code to find the optimal time $T_{opt}$ and the corresponding pipelining strategy.

\begin{algorithm}[t!]
    \caption{DP-based Pipeline Partition Strategy}
    \label{strategy}
    \small
    \begin{algorithmic}[1] % The number tells where the line numbering should start
    \REQUIRE
        \STATE $\mathbb{T}$: the transformer model with $L$ and $P$; 
         \STATE $\mathbb{D}$: the list of available devices;
         \STATE $\mathbb{B}$: bandwidth between devices;
      \ENSURE
        \STATE $T_{\text{opt}}$: the optimal time for maximum throughput;
        \STATE $\mathbb{R}$: the specific strategy for the optimal time; 
        % \PROCEDURE{Partition}{$\mathbb{T},\mathbb{D}, \mathbb{B}$}
      
        \STATE Initial $h(i, \mathbb{S}, u) \gets +\infty$ for all $i,\mathbb{S},u$;
        \STATE Initial $h(0, \emptyset, \varnothing) \gets 0$;
        \STATE Initial $answer \gets \infty$;
        \FOR{$i=0$ to $L-1$}
           \FOR{\textbf{each} subset $\mathbb{S} \subseteq \mathbb{D}$ }
                \FOR{\textbf{each} $u \in \mathbb{D} \setminus \mathbb{S}$}
                    \FOR{$j=i+1$ to $L$}
                        \IF{$\sum_{k=i}^j M_{k} > m_u$}
                            \STATE Break;
                        \ENDIF
                        
                        \STATE Calculate Eq.\eqref{transistion_equation} and assign its value to $C$;
                        \IF{$j == L$}
                            \IF{$C < answer $}
                                \STATE $answer = C$;
                                \STATE $index = (L, \mathbb{S}, u)$;
                            \ENDIF
                        \ELSE
                            \FOR{\textbf{each} $v \in \mathbb{D} \setminus \mathbb{S} \setminus \{u\}$}
                                \IF{$C < h(j, \mathbb{S} \cup \{u\}, v)$}
                                    \STATE $h(j, \mathbb{S} \cup \{u\}, v) = C$;
                                    \STATE $precursor(j, \mathbb{S} \cup \{u\}, v) = (i, u) $
                                \ENDIF
    
                            \ENDFOR
                        \ENDIF
                    \ENDFOR
                \ENDFOR
            \ENDFOR
        \ENDFOR
         
        \STATE \text{// Find the optimal results}
        \STATE Initial $T_{\text{opt}} \gets +\infty$;
        \WHILE{enumerate each subset $\mathbb{S} \subseteq \mathbb{D}$ }
                \STATE $T_{\text{opt}} = \min(h(L, \mathbb{S}, \varnothing) ,T_{\text{opt}})$
        \ENDWHILE
        
        \STATE \text{// Find the optimal strategy}
        \STATE $(i, \mathbb{S}, u) =  index$;
        \STATE Add $(i+1\rightarrow L, u) $ to $\mathbb{R}$;
        \WHILE{$i > 0$}
            \STATE $(i, u) = precursor(index)$
            \STATE Add $(i+1 \rightarrow index[0], u)$ to $\mathbb{R}$;
            \STATE $index = (i, \mathbb{S}\setminus{u}, u)$
        \ENDWHILE
        
        \STATE {\bfseries Return:} $T_{\text{opt}}, \mathbb{R}$
       % \ENDPROCEDURE
    \end{algorithmic}
\end{algorithm}

The computational complexity of the proposed algorithm is $O( 2^D \times L^2\times D^2 )$, where $D$ is the number of available devices and $L$ is the number of layers. The $2^D$ factor is due to the assumption that all devices are distinct. As a comparison, for the naive brute force solution, the search space is $\sum_{i=1}^{\min\{D, L\}}\frac{D!}{(D-i)!}\binom{L-1}{i-1} \gg D! \gg 2^D$, which has a much higher complexity. Moreover, in most scenarios, there should exist identical devices with the same computation and communication capabilities. Therefore, the number of devices $D$ could be divided into $N$ categories, where the $i$-th category has $n_i$ devices ($\sum_{i=1}^{N}n_i = D$). The search state of DP can be further reduced, hence the computation complexity could be reduced to $O(\prod_{i=1}^N(n_i+1)\times L^2\times N^2)$. For instance, consider the case where there are three types of devices, $N=3$, and each type has the same number of devices, i.e., $n_1 = n_2 = n_3 = n$.  Then the actual computational complexity is $O((n+1)^3 \times L^2 \times N^2) = O(9\times (n+1)^3 \times L^2)$. For example, given $N=3$ device types, where each type has $n=3$ devices, we measure the execution time for these three methods using the ViT-Base model on a 1.6 GHz Intel Core i5 CPU and present the results in Table~\ref{tab:algorithm_method}.

\begin{table}[t]
  \caption{Partitioning method performance.}
  \label{tab:algorithm_method}
  \small
 \begin{minipage}{\columnwidth}
  \begin{center}
  \begin{tabular}{c|c}
    \toprule
    Algorithm & Time \\ 
    \midrule
    Category dynamic programming  & 0.01 sec \\
    Naive dynamic programming   & 18.6 sec \\
    Brute force  search &  71 min   \\ \hline
  \bottomrule  
%   \begin{tabular}{p{0.3\linewidth} | p{0.5\linewidth}}
    % \toprule
    % Name &  Configuration\\
    % \midrule
    % Device name  & MinnowBoard \\
    % Processors    & Intel Atom E3845 \\
    % CPU Cores    & 4 $\times$ 1.91 GHz \\    
    % Memory size         &  2 GB\\
    % Max bandwidth        &  1 Gbps \\ 
    % \midrule
    % Device name  & RCC-VE Network Board \\
    % Processors &  Intel Atom C2558 \\
    % CPU Cores    & 4 $\times$ 2.4 GHz \\ 
    % Memory size  &  8 GB \\ 
    % Max bandwidth  &  1 Gbps \\ \hline
% \vspace{-4mm}
\end{tabular}
\vspace{-4mm}
\end{center}
\end{minipage}
\end{table}
%%%%%%%%%%%%%%%%%%%%%%%%%%%%%%% Partitioning End %%%%%%%%%%%%%%%%%%%%%%%%%%%%%%%%%%%%%%%
%%%%%%%%%%%%%%%%%%%%%%%%%%%%%%%%%%%%%%%%%%%%%%%%%%%%%%%%%%%%%%%%%%%%%%%%%%%%%%%%%%%%%%%%

\section{Experimental Setup}
\label{experimental_setup}
%%%%%%%%%%%%%%%%%%%%%%%%%%%%%%% Experiments Start %%%%%%%%%%%%%%%%%%%%%%%%%%%%%%%%%%%%%%
%%%%%%%%%%%%%%%%%%%%%%%%%%%%%%%%%%%%%%%%%%%%%%%%%%%%%%%%%%%%%%%%%%%%%%%%%%%%%%%%%%%%%%%%
This section describes the experimental hardware, software dependencies, and evaluation baselines.

\noindent\textbf{Testbed.}  We conduct experiments on the Dispersed Computing Program Testbed (DCompTB) for edge computing platforms \cite{MergeTB}. DCompTB exposes the two edge device types described in Table~\ref{tab:dcomptb}: a MinnowBoard and an RCC-VE Network Board. For evaluation, we first configure homogeneous edge clusters using both MinnowBoards and RCC-VE Network Boards, where each cluster is composed of identical devices and network bandwidth. For heterogeneous experiments, we mix device types and further increase heterogeneity by leveraging system software tools. We use the \texttt{cpulimit} tool to limit the CPU usage and the \texttt{ulimit} tool to limit the memory size on the RCC-VE boards. We then use the \texttt{tc} tool to vary network bandwidth between edge devices. In multiple edge computing scenarios, it is common to have tens of milliseconds latency \cite{latency}, thus we impose a fixed 20 ms latency when varying the bandwidth.

\begin{table}[t]
  \caption{DCompTB device configurations.}
  \label{tab:dcomptb}
  \small
  \begin{minipage}{\columnwidth}
  \begin{center}
  \begin{tabular}{c|c}
    \toprule
    Configuration &  Value\\
    \midrule
    Device name  & MinnowBoard \\
    Processors    & $4\times$ Intel Atom E3845 @ 1.91 GHz \\
    % CPU Cores    & 4 $\times$ 1.91 GHz \\    
    Memory size         &  2 GB\\
    Max bandwidth        &  1 Gbps \\ 
    \midrule
    Device name  & RCC-VE Network Board \\
    Processors &  $4\times$ Intel Atom C2558 @ 2.4 GHz \\
    % CPU Cores    & 4 $\times$ 2.4 GHz \\ 
    Memory size  &  8 GB \\ 
    Max bandwidth  &  1 Gbps \\ \hline
  \bottomrule
\end{tabular}
\vspace{-4mm}
\end{center}
\end{minipage}
\end{table}

\noindent\textbf{Software.} Each node runs Debian GNU/Linux 10 with Linux kernel 4.19.0-11-amd64. We use PyTorch 1.9.0 as the deep learning inference engine and the PyTorch RPC library with the TensorPipe backend as the distributed framework \cite{NEURIPS2019_9015}.

\noindent\textbf{Deep Learning Models.} The definition, configuration, and implementation of the ViT-Base, ViT-Large, and ViT-Huge models are taken from HuggingFace 4.10.0 \cite{wolf-etal-2020-transformers}. We evaluate ViT models for the image classification task. Input images are from ImageNet 2012 \cite{deng2009imagenet} and resized after the embedding layer with a uniform input dimension to the transformer layers. 

\noindent\textbf{Baselines.} To the best of our knowledge, this is the first work that evaluates pipeline parallelism for inference on heterogeneous edge devices, so we adopt baselines from pipelines developed for training in data centers. We re-implement GPipe \cite{huang2019gpipe} with an even partitioning method for inference on the CPU as the baseline. Although PipeDream targets asynchronous parallel training, its partition method can be applied to inference by considering only the forward pass. PipeDream also provides open-source code for partitioning for inference \cite{pipedream}. PipeDream considers a hierarchical interconnect that represents a data center interconnect topology, but that does not model the ad hoc networks of edge systems. To apply PipeDream to the edge, we assume a one-level communication network and compare with its pipeline partitioning scheme. We choose the optimal microbatch size for GPipe and PipeDream and compare the system performance.

%%%%%%%%%%%%%%%%%%%%%%%%%%%%%%% Experiments End %%%%%%%%%%%%%%%%%%%%%%%%%%%%%%%%%%%%%%%%
%%%%%%%%%%%%%%%%%%%%%%%%%%%%%%%%%%%%%%%%%%%%%%%%%%%%%%%%%%%%%%%%%%%%%%%%%%%%%%%%%%%%%%%%
\begin{figure*}[!th]
    \centering
    \includegraphics[width=0.9\linewidth]{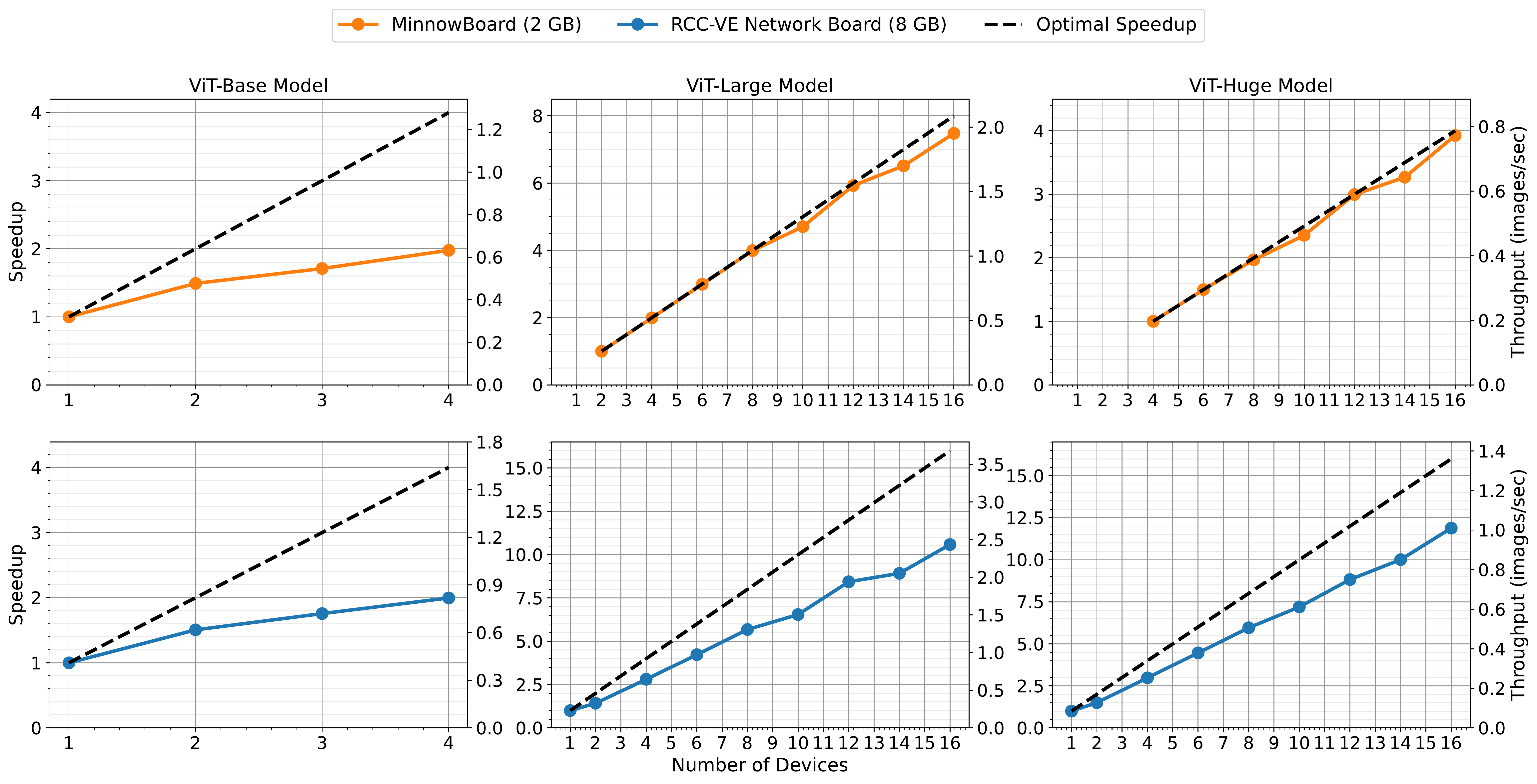}
    \vspace{-4mm}
    \caption{EdgePipe's throughput performance for three ViT  models on two homogeneous edge clusters. The ViT-Large and ViT-Huge models cannot fit on a single MinnowBoard.}
    \label{fig:exp_1_performance}
\end{figure*}
\section{Evaluation}
\label{evaluation}
%%%%%%%%%%%%%%%%%%%%%%%%%%%%%% Evaluation start %%%%%%%%%%%%%%%%%%%%%%%%%%%%%%%%%%%%%%%%
%%%%%%%%%%%%%%%%%%%%%%%%%%%%%%%%%%%%%%%%%%%%%%%%%%%%%%%%%%%%%%%%%%%%%%%%%%%%%%%%%%%%%%%%

This section describes the EdgePipe experimental evaluation.  We first show the runtime performance on homogeneous clusters using two DCompTB device types.  We then demonstrate the effectiveness of our partitioning method on heterogeneous clusters. We also explore the effects of communication bandwidth and the relationship between microbatch size and throughput. Finally, we evaluate EdgePipe with compressed models.

\subsection{Runtime Performance Analysis} 

%% Demonstrate our results: how good are we,
We first evaluate EdgePipe's performance on the 2 GB MinnowBoard and the 8 GB RCC-VE Network Board devices in homogeneous clusters.  Figure~\ref{fig:exp_1_performance} presents throughput on these clusters for up to 16 stages (devices).

For MinnowBoard devices, we achieve $0.63$ images per second throughput with $4$ devices using the ViT-Base model, which is $1.98\times$ faster compared to the single-device performance.
The ViT-Large and ViT-Huge models cannot fit in memory on a single MinnowBoard device. Hence, we use 2-stage and 4-stage throughput as the speedup baselines for the ViT-Large and ViT-Huge models, respectively. With 16 MinnowBoard devices, EdgePipe achieves $1.95$ images per second throughput, which is a $7.48\times$ speedup over the 2-stage baseline (where the optimal speedup is 16/2=8). For the ViT-Huge model, EdgePipe achieves $0.77$ images per second throughput with $16$ MinnowBoard devices, which gives a $3.93\times$ speedup over the 4-stage baseline (where the optimal speedup is 16/4=4).

We achieve similar scalability on the RCC-VE Network Board devices. With the ViT-Base model, EdgePipe achieves $0.82$ images per second throughput with $1.99\times$ speedup with four devices. Compared with the single-device baseline, EdgePipe achieves $2.43$ and $1.01$ images per second throughput for the ViT-Large and ViT-Huge models with $10.59\times$ and $11.88\times$ speedup using $16$ devices, respectively.

\begin{figure}[t]
    \centering
    \includegraphics[width=0.75\linewidth]{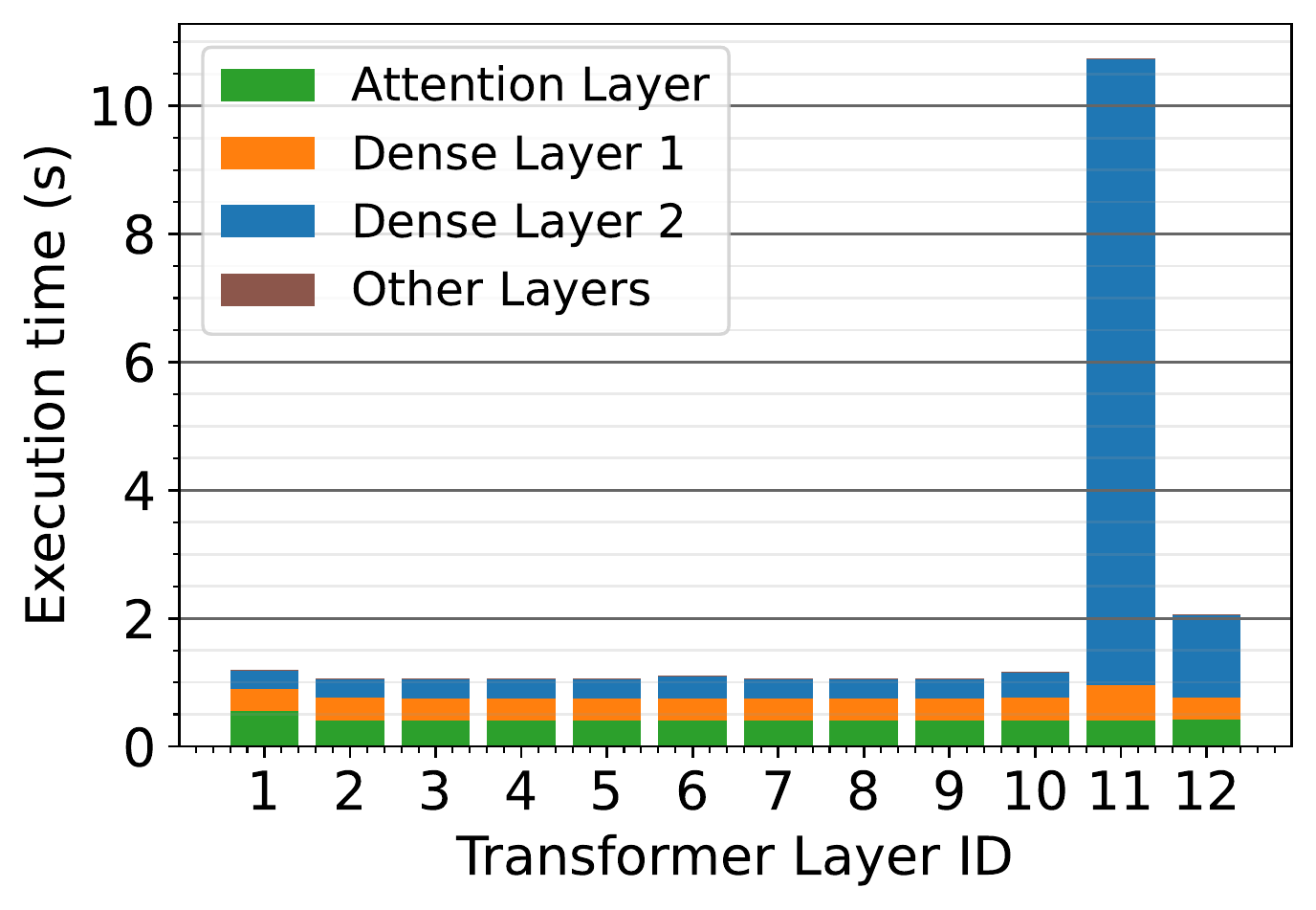}
    \vspace{-4mm}
    \caption{ViT-Base model layer execution times on a MinnowBoard with batch size $8$.}
    \vspace{-4mm}
    \label{fig:layer_level_execution_time}
\end{figure}

The sub-linear performance for the ViT-Base model is primarily due to uneven execution times of different stages. We observe the performance difference for the slowest and fastest stages is  about $25\%$ for the 2-stage pipeline. With $4$ stages, this time difference increases to $80\%$ and causes a more serious performance loss. Figure~\ref{fig:layer_level_execution_time} illustrates the issue by quantifying each ViT-Base layer's execution time on the MinnowBoard. The second dense layer in the $11$-th transformer layer, which only includes the linear transformation with the matrix multiplication operation and thus cannot be further partitioned with pipeline parallelism, requires considerably more execution time than other layers. The variation in execution time is due to differences in the sparsity of weights.  We observe the similar performance behavior on the RCC-VE Network Board devices. In contrast, each transformer layer in the ViT-Large and ViT-Huge models have similar inference times, so EdgePipe scales better on the ViT-Large and ViT-Huge models. In addition, we observe compressed models reduce stage imbalances and mitigate this challenge. We report on compressed models in subsection \ref{model_compression}.

EdgePipe achieves nearly linear performance improvements with the ViT-Large and ViT-Huge models when pipelining with both edge device types.  EdgePipe achieves $3.93\times$ speedup (over a 4-node baseline) on 16 MinnowBoard devices and $11.88 \times$ speedup on 16 RCC-VE Network Board devices with the ViT-Huge model. These results demonstrate EdgePipe's effectiveness for large-scale models, including ones that otherwise cannot fit on single devices.

%%%%%% Heterogeneous Clusters %%%%%%%%%%%%
\subsection{Heterogeneous Clusters}
\label{heterogenous_clusters}

\begin{figure*}[t]
    \centering
    \includegraphics[width=0.9\linewidth]{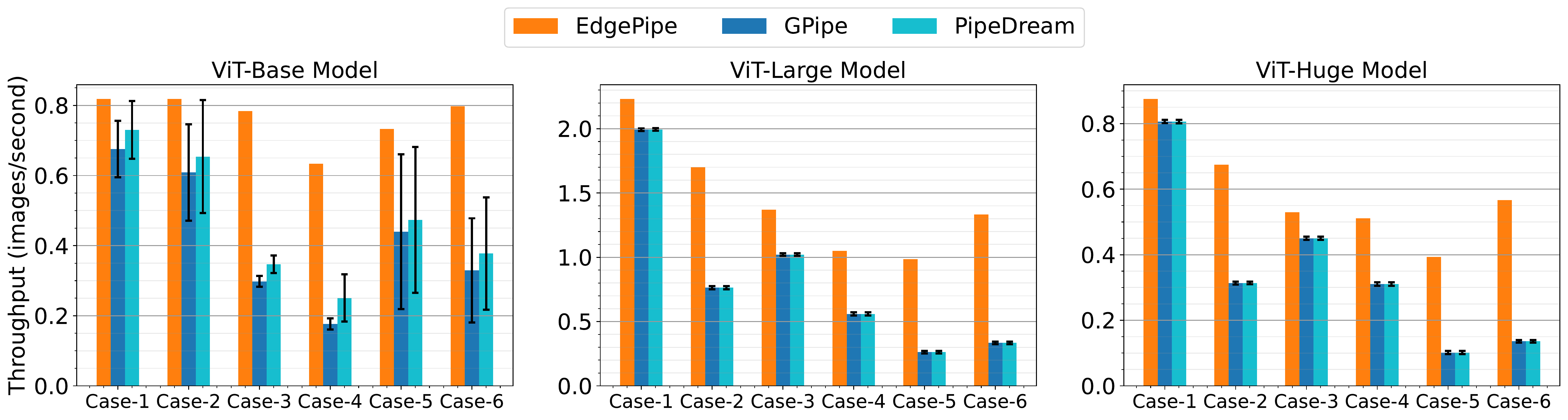}
    \vspace{-4mm}
    \caption{Throughput results for heterogeneous clusters with six cases.  We report the average throughput and mark the variance with 10 random device orders for GPipe and PipeDream. GPipe and PipeDream show identical partitioning results for the ViT-Large and ViT-Huge models.}
    \vspace{-4mm}
    \label{fig:exp_7_heter}
\end{figure*}

Compared with the data center, edge devices are more heterogeneous in computation and communication capabilities.  To better simulate heterogeneous resource-constrained devices, we throttle the CPU usage of the RCC-VE Network Boards to emulate diminished inference performance. We also vary the maximum bandwidth for both MinnowBoard and RCC-VE Network Board devices to emulate different network link capacities. We compare EdgePipe with GPipe and PipeDream on six clusters with increasing heterogeneity.  Because GPipe and PipeDream do not specify the device mapping order, we test them with $10$ random device orders and measure average performance and variance.  Device configuration details are presented in Table~\ref{tab:case}, and experimental results are shown in Figure~\ref{fig:exp_7_heter}.

% Figure~\ref{fig:heter_single_performance} shows RCC-VE Network Board performance correlated to different CPU utilizations.   The experimental results are shown in Figure~\ref{fig:exp_7_heter}.
% We use six heterogeneous cluster configurations with increasingly heterogeneity to verify the effectiveness of our pipeline partitioning strategy.  Device configuration details are presented in Table~\ref{tab:case}. 
% To better simulate heterogeneous resource-constrained devices, we throttle the CPU usage of the RCC-VE Network Boards to emulate diminished inference performance.  We also vary the maximum bandwidth for both MinnowBoard and RCC-VE Network Board devices to emulate different network link capacities.
% We use six heterogeneous cluster configurations with increasingly heterogeneity to verify the effectiveness of our pipeline partitioning strategy.  Device configuration details are presented in Table~\ref{tab:case}.  

\begin{table}[t]
  \caption{Heterogeneous cluster configurations.}
  \label{tab:case}
  \scriptsize
%   \begin{minipage}{2\columnwidth}
  \begin{center}
  \begin{tabular}{ccccc}
    \toprule
% Case \#     &  Device A & Device B \\
  Case & Devices  & CPU  & Memory & Bandwidth  \\
      \midrule
    \multirow{2}{*}{1}     &  $8\times$RCC-VE & 100\% & 8 GB &  1 Gbps \\
    &$8\times$MinnowBoard & 100\% & 2 GB & 1 Gbps \\ \hline
    \multirow{4}{*}{2} & $4\times$RCC-VE & 100\% & 8 GB &  1 Gbps\\
    & $4\times$RCC-VE & 75\% & 4 GB & 1 Gbps\\
    &$4\times$RCC-VE & 25\% & 4 GB & 1 Gbps \\
    & $4\times$MinnowBoard & 100\% & 2 GB & 1 Gbps\\\hline
    \multirow{2}{*}{3}     &  $8\times$RCC-VE & 100\% & 8 GB & 40 Mbps \\
    & $8\times$MinnowBoard & 100\% & 2 GB & 10 Mbps \\\hline
    
     \multirow{4}{*}{4} &  $4\times$RCC-VE & 100\% & 8 GB & 30 Mbps \\
     & $4\times$RCC-VE & 100\% & 8 GB & 20 Mbps\\
    & $4\times$MinnowBoard & 100\% & 2 GB & 10 Mbps\\
    & $4\times$MinnowBoard  & 100\% & 2 GB & 5 Mbps\\\hline
    
     \multirow{3}{*}{5} &  $3\times$RCC-VE & 100\% & 8 GB & 50 Mbps\\
     &$8\times$RCC-VE & 10\% & 4 GB & 20 Mbps\\
     & $5\times$MinnowBoard & 100\% & 2 GB & 30 Mbps \\\hline
    
     \multirow{6}{*}{6} &  $2\times$RCC-VE & 100\% &  8 GB & 100 Mbps \\
     &$3\times$RCC-VE & 75\% & 4 GB &  60 Mbps \\
     & $4\times$RCC-VE  & 50\% &  4 GB & 40 Mbps\\
     & $3\times$RCC-VE & 25\% & 4 GB &  20 Mbps \\
     & $2\times$RCC-VE & 10\% & 4 GB & 10 Mbps \\
     & $2\times$MinnowBoard & 100\% & 2 GB & 80 Mbps\\ \hline       \hline
  \end{tabular}
  \vspace{-5mm}
\end{center}
% \end{minipage}
\end{table}
% \begin{figure}[t]
%     \centering
%     \includegraphics[width=0.8\linewidth]{mlsys2022style/figures/Experiment_6_single_devices.pdf}
%     \caption{RCC-VE Network Board performance with CPU utilization limits.}
%     \label{fig:heter_single_performance}
% \end{figure}

By comparing Cases~1 and~2, we show the effect of introducing heterogeneous computing capabilities on system performance. For the ViT-Base model, EdgePipe achieves the best performance of $0.82$ images per second in both cases.  GPipe and PipeDream show a significant variance with different device orders for the ViT-Base model.  In Case~1, GPipe's throughput ranges from $0.57$ to $0.76$ images per second, and PipeDream's throughput ranges from $0.64$ to $0.82$ images per second.  In Case~2, which introduces more heterogeneity, GPipe and PipeDream show a larger variance of $0.47$ to $0.75$ and $0.50$ to $0.82$ for the ViT-Base model. For the ViT-Large and ViT-Huge models, both GPipe and PipeDream adopt the even partitioning strategy and obtain the same performance. EdgePipe achieves $2.23$ and $1.69$ images per second for the ViT-Large model and  $0.88$ and $0.67$ images per second for the ViT-Huge model in Cases~1 and~2, respectively. For the same two cases, GPipe and PipeDream only achieve $1.99$ and $0.76$ images per second for the ViT-Large model and $0.81$ and $0.31$ images per second for the ViT-Huge model. EdgePipe demonstrates both better performance and robustness when compute capabilities are more heterogeneous.

% CKI: Moved this up here so it would appear on same page as text description
\begin{figure*}[t]
    \centering
    \includegraphics[width=0.85\linewidth]{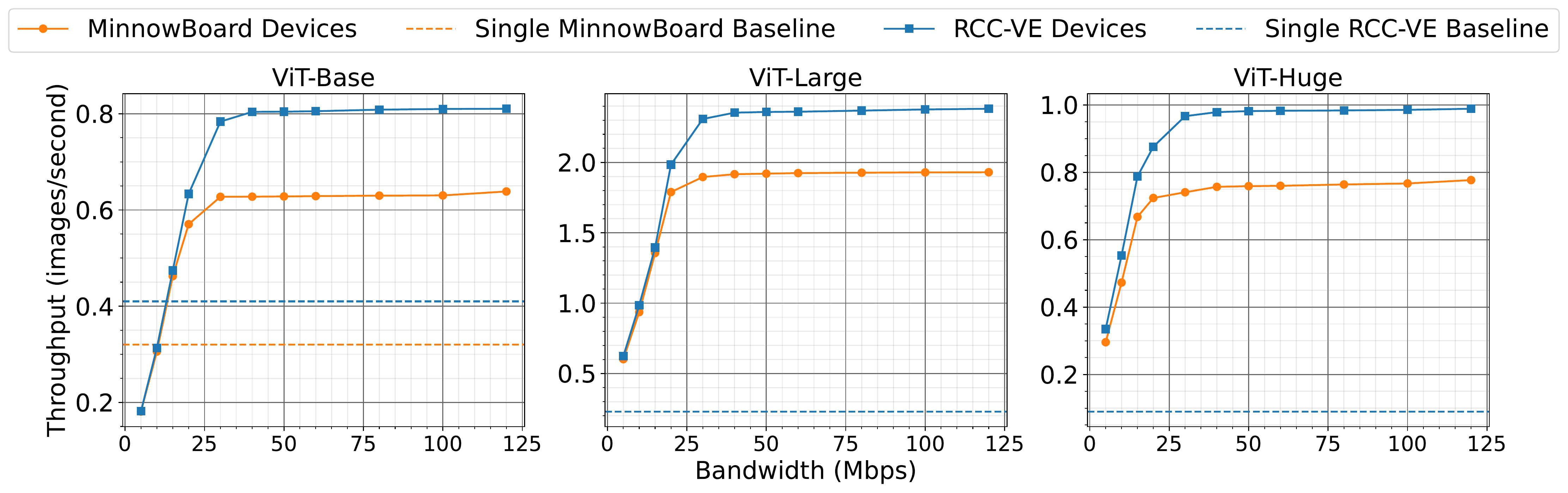}
    \vspace{-4mm}
    \caption{The relationship between bandwidth and system throughput.}
    \vspace{-2mm}
    \label{fig:exp_1_bandwidth}
\end{figure*}

Case~3 has the same compute resources as Case~1, but with less communication bandwidth. We further vary the bandwidth in Case~4 with the same compute resources.  For the ViT-Base model, EdgePipe achieves the best throughput of $0.78$ and $0.63$ images per second in both Cases~3 and~4. GPipe and PipeDream show a significant performance degradation due to the limited bandwidth. In Case~3 for the ViT-Base model,  GPipe's throughput ranges from $0.29$ to $0.32$ images per second, and PipeDream's throughput ranges from $0.32$ to $0.38$ images per second. In Case~4 for the same model,  GPipe's throughput ranges from $0.16$ to $0.20$ images per second, and PipeDream's throughput ranges from $0.18$ to $0.32$ images per second. 
% EdgePipe achieves $2.05\times$  and $2.51\times$ speedup compared to the average throughput in Case~3 and Case~4, respectively. 
For the ViT-Large and ViT-Huge models, EdgePipe achieves the best performance with fewer devices than GPipe and PipeDream for Cases~3 and~4. For the ViT-Large and ViT-Huge models on Case~3, EdgePipe selects $8$ devices with $40$ Mbps bandwidth and one device with $10$ Mbps bandwidth as the last stage and achieves throughput of $1.37$ and $0.53$ images per second, respectively. GPipe and PipeDream use $16$ devices and performance degrades with $1.02$ and $0.44$ images per second for the ViT-Large and ViT-Huge models in Case~3.  In Case~4, EdgePipe selects $7$ devices for the ViT-Large model and $9$ devices for the ViT-Huge model to achieve throughput of $1.05$ and $0.51$ images per second, respectively. GPipe and PipeDream achieve $0.55$ and $0.31$ for the ViT-Large and ViT-Huge models.  In Case~4, EdgePipe shows $1.90\times$ and $1.54\times$ speedup compared to PipeDream for the ViT-Large and ViT-Huge models, respectively. These two cases demonstrate the effectiveness of EdgePipe's partitioning strategy for heterogeneous network.

In Cases~5 and~6, we mix the heterogeneity of devices and networks. In Case~5, we added $8$ extremely resource-constrained devices with CPUs at $10\%$ capacity and $20$ Mbps bandwidth. EdgePipe achieves the best throughput with $0.73$, $0.99$, and $0.39$ images per second using $4$, $7$, and $7$ devices for the ViT-Base, ViT-Large, and ViT-Huge models. For the ViT-Base model in Case~5, GPipe's throughput ranges from $0.22$ to $0.66$ images per second, and PipeDream's throughput ranges from $0.26$ to $0.69$ images per second. For the ViT-Large and ViT-Huge models in Case~5, GPipe and PipeDream achieve $0.26$ and $0.10$ images per second.   In Case~5, EdgePipe shows $1.55\times$, $3.75\times$, $3.84\times$ speedup relative to PipeDream's average throughput for the ViT-Base, ViT-Large, and ViT-Huge models, respectively.  Case~6 shows a scenario with $6$ types of devices weighted toward devices with medium performance. In this case, EdgePipe uses $4$, $12$, and $14$ devices to achieve $0.80$, $1.33$, and $0.57$ images per second for these three ViT models. For the ViT-Base model in Case~6, GPipe's throughput ranges from $0.18$ to $0.48$, and PipeDream's throughput ranges from $0.22$ to $0.54$ images per second. GPipe and PipeDream achieve $0.33$ and $0.14$ images per second for the ViT-Large and ViT-Huge models, respectively. EdgePipe achieves speedup of $1.98\times$, $3.98\times$, and $4.16\times$ for the ViT-Base, ViT-Large, and  ViT-Huge models compared to PipeDream's average throughput.   Cases 5 and 6 demonstrate EdgePipe's ability to  schedule around low-performance devices and map the task reasonably to achieve the best throughput. 
% Both Case~5 and~6  demonstrate that EdgePipe can provide a reasonable and efficient partition strategy for the heterogeneous cluster to achieve the better performance.  

EdgePipe performs significantly better than the GPipe and PipeDream partition methods on all six heterogeneous clusters. Unlike GPipe and PipeDream, EdgePipe successfully avoids the lowest-performing devices by considering multiple factors in exploiting pipelining to improve performance.

%%%%%%%%%%%%%%%%%%%%%%%%%%%%%%%%%%%%%%%%%%%%%%%%%%%

%%%%%%%%%%%%%%%% Bandwidth %%%%%%%%%%%%%%%%%%%
\subsection{Impact of Bandwidth}

Edge computing often has more limited bandwidth between devices compared to the data center. To evaluate the relationship between system performance and bandwidth, we vary the bandwidth between all devices from 120 Mbps to 5 Mbps. We test with 4 pipeline stages using the ViT-Base model and 16 pipeline stages using the ViT-Large and ViT-Huge models.
As shown in Figure~\ref{fig:exp_1_bandwidth}, performance does not decrease significantly until bandwidth drops below 30 Mbps for all three ViT models, demonstrating the feasibility of our system for practical edge applications.
Upon further reduction in bandwidth from 30 Mbps to 5 Mbps, although the performance shows a linear decline, the throughput still shows some improvements compared to the single-device baseline. For the ViT-Base model with $15$ Mbps bandwidth,  MinnowBoards and RCC-VE Network Boards show about $1.48\times$  and $1.15\times$ speedup compared to the single-device baseline. For the ViT-Large and ViT-Huge models with 5 Mbps, EdgePipe still achieves $2.69\times$ and $3.67\times$ speedup compared to the single device baseline on RCC-VE Network Board devices.
These results demonstrate that EdgePipe is still effective for large-scale models under limited network conditions. 
%%%%%%%%%%%%%%%%%%%%%%%%%%%%%%%%%%%%%%%%%%%%%%%%%%%
%%%%%%%%%%%%%%%%%%%%% Microbatch Size %%%%%%%%%%%%%
\subsection{Impact of Microbatch Size} 
\begin{figure}[!t]
    \centering
    \includegraphics[width=0.75\linewidth]{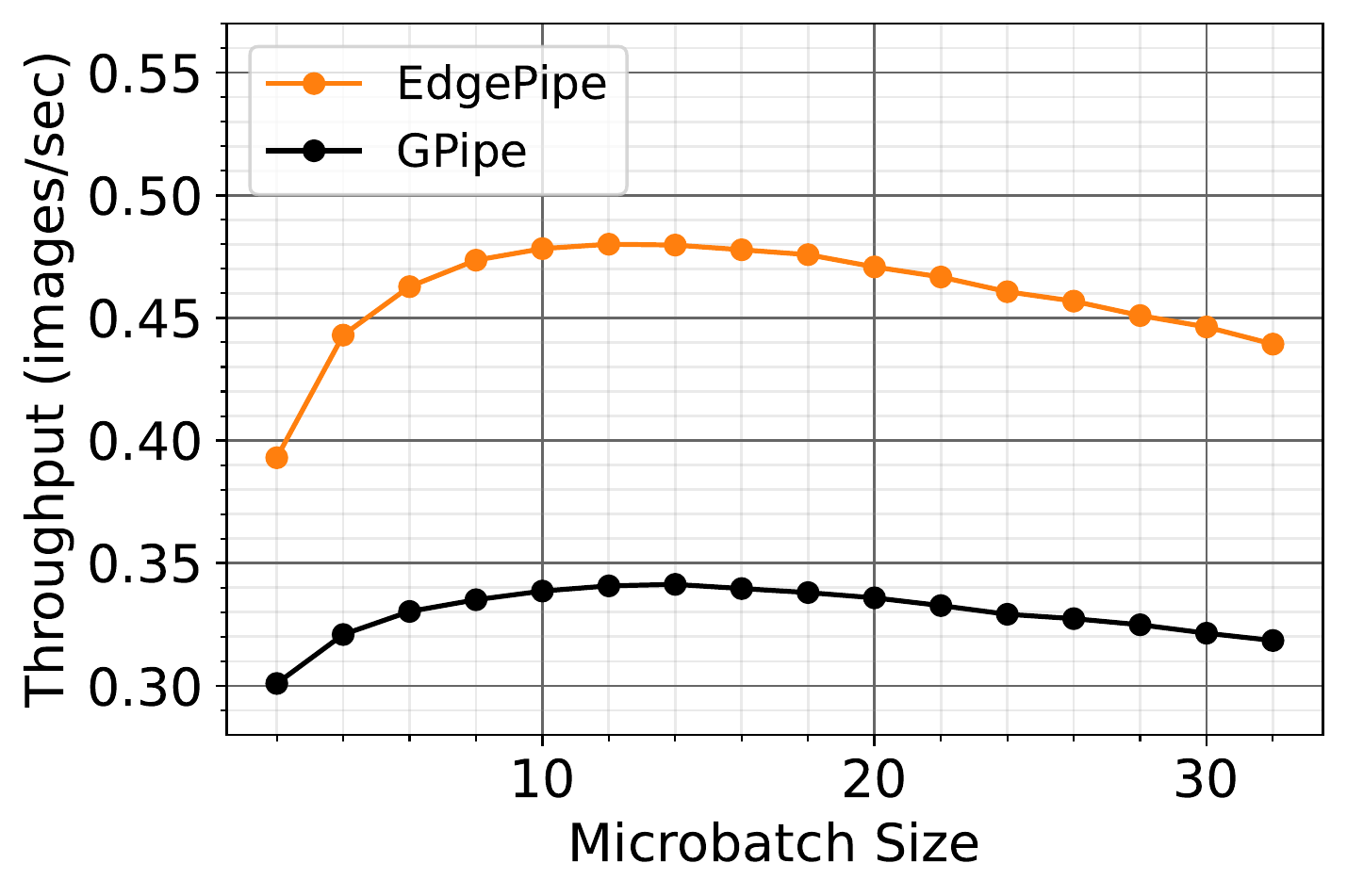}
    \vspace{-4mm}
    \caption{ Microbatch size vs. throughput relationship for ViT-Base on MinnowBoard. The model is partitioned into $2$ stages.}
    \vspace{-4mm}
    \label{fig:exp2_microbatch_4_stages}
\end{figure}

As in other pipeline frameworks, performance is affected by microbatch size. We demonstrate the relationship between throughput and microbatch size in Figure~\ref{fig:exp2_microbatch_4_stages}. For a $2$-stage pipeline, the maximum throughput of the even partitioning method in GPipe is around $0.34$ images per second with a microbatch size $12$. The throughput shows a significant increase before microbatch size $14$, which is mainly due to the continuous increase in CPU utilization as the microbatch size increases. Beyond this size, the throughput begins to slowly decline because larger microbatch sizes reduce the efficiency of the pipeline parallelism. There is a similar pattern for microbatch size and the system throughput for EdgePipe. However, it achieves a significant performance improvement with a maximum throughput of $0.48$ images per second with a microbatch size of $12$. The fine-grained partitioning method in EdgePipe achieves more efficient CPU utilization than even partitioning. Other pipeline stages show similar behavior. To give a fair comparison of throughput, we choose the optimal microbatch and its related performance for all methods in the evaluations.

\subsection{EdgePipe with Model Compression}
\label{model_compression}
\begin{figure}[t]
    \centering
    \includegraphics[width=0.75\linewidth]{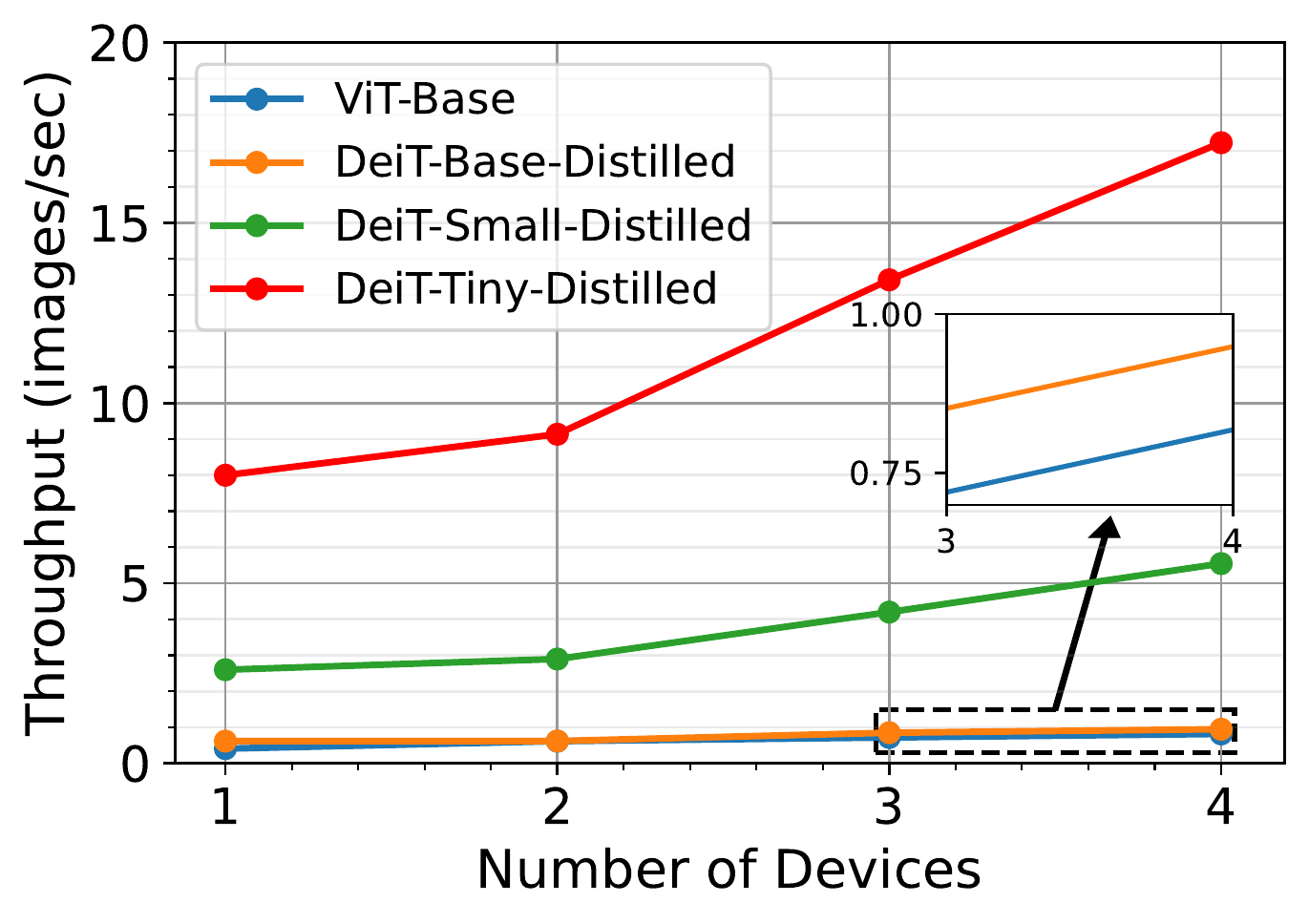}
    % \vspace{-4mm}
    \caption{The throughput of DeiT distilled models compared to the ViT-Base model on RCC-VE Network Boards.}
    % \vspace{-4mm}
    \label{fig:deit}
\end{figure}

% the importance of model compression techniques; shrink the model size and achieve better throughput, which is an important comp to EdgePipe
Model compression techniques shrink the model size to reduce compute cost and potentially accelerate inference \cite{han2015deep,hinton2015distilling} and training \cite{kundu2020pre}. These approaches can be considered as important complementary strategies to EdgePipe to improve the performance on resource-constrained platforms. For example, compared to the ViT-Base model, DeiT-Tiny and DeiT-Small use distillation to achieve similar ImageNet top-1 accuracy with compressed models of up to $17.2\times$ and $3.9\times$, respectively \cite{deit2021}. To demonstrate the efficacy of using compressed models with EdgePipe, we evaluate DeiT-Base, Small, and Tiny on up to $4$ RCC-VE Network Board devices. We also provide the throughput of the ViT-Base model on same devices for comparison. Figure~\ref{fig:deit} presents the throughput results.

DeiT-Base, which has identical model structure as ViT-Base, achieves $0.62$ images per second throughput on a single RCC-VE Network Board. With $4$ devices, the DeiT-Base model achieves $0.95$ images per second and outperforms ViT-Base's $0.82$ images per second. DeiT-Small and DeiT-Tiny demonstrate more significant improvements. In particular, DeiT-Small and DeiT-Tiny can provide throughput of up to $5.55$ and $17.23$ images per second, respectively. Interestingly, we observe DeiT-Small and DeiT-Tiny models ease the uneven execution times seen with ViT-Base and demonstrate better scalability with EdgePipe. The model compression technique, as an orthogonal method, can potentially further improve the performance of EdgePipe, making this combined approach a promising solution for large scale model inference at the edge.

% \subsection{Analysis}

% \textbf{Generalize to other models}

% \textbf{Model Compression friendly}

%%%%%%%%%%%%%%%%%%%%%%%%%%%%%%% Evaluation End %%%%%%%%%%%%%%%%%%%%%%%%%%%%%%%%%%%%%%%%%
%%%%%%%%%%%%%%%%%%%%%%%%%%%%%%%%%%%%%%%%%%%%%%%%%%%%%%%%%%%%%%%%%%%%%%%%%%%%%%%%%%%%%%%%

\section{Related Work}
\label{related_work}
%%%%%%%%%%%%%%%%%%%%%%%%%%%%%% Related Work Start %%%%%%%%%%%%%%%%%%%%%%%%%%%%%%%%%%%%%%
%%%%%%%%%%%%%%%%%%%%%%%%%%%%%%%%%%%%%%%%%%%%%%%%%%%%%%%%%%%%%%%%%%%%%%%%%%%%%%%%%%%%%%%%

Current techniques to enable the execution of large models on edge devices mainly fall into two categories: single device optimization and distributed inference on multiple devices or servers. Aggressive model compression is an example of single device optimization methods. EdgeBERT \cite{tambe2020edgebert} combines network pruning, entropy-based early exit, and adaptive attention span to reduce the model size and the inference latency of Bidirectional Encoder Representations from Transformers (BERT) for NLP tasks. Lite Transformer \cite{wu2019lite} adopts adaptive inference to reduce inference computation cost. Another promising solution includes neural architecture search (NAS), that trains a flexible supernet model to yield various subnets suitable for different targeted hardware platforms \cite{hat}. However, most of the above methods are either not suitable for distributed platform or need redesigning and retraining of a pre-trained models and can potentially incur non-negligible drop in accuracy.

Through the assistance of cloud servers or distributed edge devices, the latency and computations for each device can be reduced without sacrificing accuracy. In several works \cite{DNNOff,Neurosurgeon,JointDNN}, the distributed inference of DNN models on edge devices is partitioned and offloaded to cloud servers to reduce the minimize the latency and computations. Considering the limited bandwidth and uncertain delay between the edge and the cloud, MoDNN \cite{mao2017modnn} employs a MapReduce-like distributed inference paradigm and only utilizes idle mobile devices to execute CNN models. DeepThings \cite{zhao2018deepthings} proposes a fine-grain partition method for CNN models on edge clusters. In \cite{AdaptiveExecution}, the proposed adaptive parallel inference method for CNN models  is extend to heterogeneous edge devices. 

With the emergence of transformer-based models, the model size continues to increase making distributed execution more important. Megatron-LM \cite{Megatron-LM2019} implements operation partitioning for transformer-based models.  Pipeline parallelism is proposed to address the problem of communication overheads. GPipe \cite{huang2019gpipe} presents effective pipeline parallelism for training large models on multiple TPU accelerators. PipeDream \cite{pipedream} and its subsequent work \cite{pipedream_2bw} target  heterogeneous platforms and adopt pipeline parallelism to accelerate training. PipeMare \cite{pipemare} proposes a memory-efficient  pipeline parallelism without sacrificing utilization. These work target data centers, and are difficult to directly apply to edge computing.
%%%%%%%%%%%%%%%%%%%%%%%%%%%%%% Related Work End %%%%%%%%%%%%%%%%%%%%%%%%%%%%%%%%%%%%%%%%
%%%%%%%%%%%%%%%%%%%%%%%%%%%%%%%%%%%%%%%%%%%%%%%%%%%%%%%%%%%%%%%%%%%%%%%%%%%%%%%%%%%%%%%%

\section{Conclusion}
\label{conclusion}
%%%%%%%%%%%%%%%%%%%%%%%%%%%%%% Conclusion End %%%%%%%%%%%%%%%%%%%%%%%%%%%%%%%%%%%%%%%%%%
%%%%%%%%%%%%%%%%%%%%%%%%%%%%%%%%%%%%%%%%%%%%%%%%%%%%%%%%%%%%%%%%%%%%%%%%%%%%%%%%%%%%%%%%
In this paper, we presented EdgePipe, a distributed inference acceleration system using pipeline parallelism. Unlike current pipeline parallelism frameworks for model training on cloud servers, EdgePipe focuses on heterogeneous resource-constrained devices. To address the workload balance problem for heterogeneous clusters, we design a dynamic programming-based partition method. We achieve $10.6\times$ and $11.9\times$ throughput speedup with 16 devices for the ViT-Large and ViT-Huge models, and demonstrate the ability to accelerate large-scale models on devices without sufficient memory. EdgePipe demonstrates effectiveness and robustness for multiple heterogeneous cases, e.g., we show up to $4.16\times$ throughput speedup compared to GPipe and PipeDream when using a heterogeneous set of devices. Finally, we demonstrate the efficacy of our proposed scheme on compressed models to show that the throughput benefits of our pipelining approach are complementary to the improvement from compression and that we still achieve speedup by leveraging multiple devices.  
%%%%%%%%%%%%%%%%%%%%%%%%%%%%%% Conclusion End %%%%%%%%%%%%%%%%%%%%%%%%%%%%%%%%%%%%%%%%%%
%%%%%%%%%%%%%%%%%%%%%%%%%%%%%%%%%%%%%%%%%%%%%%%%%%%%%%%%%%%%%%%%%%%%%%%%%%%%%%%%%%%%%%%%
\clearpage
\bibliographystyle{named}
\bibliography{ijcai19}

\begin{thebibliography}{}

\bibitem[\protect\citeauthoryear{Carion \bgroup \em et al.\egroup
  }{2020}]{carion2020end}
Nicolas Carion, Francisco Massa, Gabriel Synnaeve, Nicolas Usunier, Alexander
  Kirillov, and Sergey Zagoruyko.
\newblock End-to-end object detection with transformers.
\newblock In {\em European Conference on Computer Vision}, pages 213--229.
  Springer, 2020.

\bibitem[\protect\citeauthoryear{Chen \bgroup \em et al.\egroup
  }{2021}]{DNNOff}
Xing Chen, Ming Li, Hao Zhong, Yun Ma, and Ching-Hsien Hsu.
\newblock Dnnoff: Offloading dnn-based intelligent iot applications in mobile
  edge computing.
\newblock {\em IEEE Transactions on Industrial Informatics}, 2021.

\bibitem[\protect\citeauthoryear{Cheng and Kunz}{2009}]{smart_home_survey}
Jin Cheng and Thomas Kunz.
\newblock A survey on smart home networking.
\newblock {\em Carleton University, Systems and Computer Engineering, Technical
  Report SCE-09-10}, 2009.

\bibitem[\protect\citeauthoryear{Chin \bgroup \em et al.\egroup
  }{2020}]{chin2020towards}
Ting-Wu Chin, Ruizhou Ding, Cha Zhang, and Diana Marculescu.
\newblock Towards efficient model compression via learned global ranking.
\newblock In {\em Proceedings of the IEEE/CVF Conference on Computer Vision and
  Pattern Recognition}, pages 1518--1528, 2020.

\bibitem[\protect\citeauthoryear{d'Ascoli \bgroup \em et al.\egroup
  }{2021}]{d2021convit}
St{\'e}phane d'Ascoli, Hugo Touvron, Matthew Leavitt, Ari Morcos, Giulio
  Biroli, and Levent Sagun.
\newblock Convit: Improving vision transformers with soft convolutional
  inductive biases.
\newblock {\em arXiv preprint arXiv:2103.10697}, 2021.

\bibitem[\protect\citeauthoryear{Deng \bgroup \em et al.\egroup
  }{2009}]{deng2009imagenet}
Jia Deng, Wei Dong, Richard Socher, Li-Jia Li, Kai Li, and Li~Fei-Fei.
\newblock Imagenet: A large-scale hierarchical image database.
\newblock In {\em 2009 IEEE conference on computer vision and pattern
  recognition}, pages 248--255. Ieee, 2009.

\bibitem[\protect\citeauthoryear{Dosovitskiy \bgroup \em et al.\egroup
  }{2020}]{dosovitskiy2020image}
Alexey Dosovitskiy, Lucas Beyer, Alexander Kolesnikov, Dirk Weissenborn,
  Xiaohua Zhai, Thomas Unterthiner, Mostafa Dehghani, Matthias Minderer, Georg
  Heigold, Sylvain Gelly, et~al.
\newblock An image is worth 16x16 words: Transformers for image recognition at
  scale.
\newblock {\em arXiv preprint arXiv:2010.11929}, 2020.

\bibitem[\protect\citeauthoryear{Eshratifar \bgroup \em et al.\egroup
  }{2021}]{JointDNN}
Amir~Erfan Eshratifar, Mohammad~Saeed Abrishami, and Massoud Pedram.
\newblock Jointdnn: An efficient training and inference engine for intelligent
  mobile cloud computing services.
\newblock {\em IEEE Transactions on Mobile Computing}, 20(2):565--576, 2021.

\bibitem[\protect\citeauthoryear{Goodfellow \bgroup \em et al.\egroup
  }{2019}]{MergeTB}
Ryan Goodfellow, Stephen Schwab, Erik Kline, Lincoln Thurlow, and Geoff Lawler.
\newblock The dcomp testbed.
\newblock In {\em 12th {USENIX} Workshop on Cyber Security Experimentation and
  Test ({CSET} 19)}, Santa Clara, CA, August 2019. {USENIX} Association.

\bibitem[\protect\citeauthoryear{Han \bgroup \em et al.\egroup
  }{2015}]{han2015deep}
Song Han, Huizi Mao, and William~J Dally.
\newblock Deep compression: Compressing deep neural networks with pruning,
  trained quantization and huffman coding.
\newblock {\em arXiv preprint arXiv:1510.00149}, 2015.

\bibitem[\protect\citeauthoryear{He \bgroup \em et al.\egroup
  }{2021}]{he2021pipetransformer}
Chaoyang He, Shen Li, Mahdi Soltanolkotabi, and Salman Avestimehr.
\newblock Pipetransformer: Automated elastic pipelining for distributed
  training of transformers.
\newblock {\em arXiv preprint arXiv:2102.03161}, 2021.

\bibitem[\protect\citeauthoryear{Hinton \bgroup \em et al.\egroup
  }{2015}]{hinton2015distilling}
Geoffrey Hinton, Oriol Vinyals, and Jeff Dean.
\newblock Distilling the knowledge in a neural network.
\newblock {\em arXiv preprint arXiv:1503.02531}, 2015.

\bibitem[\protect\citeauthoryear{Huang \bgroup \em et al.\egroup
  }{2019}]{huang2019gpipe}
Yanping Huang, Youlong Cheng, Ankur Bapna, Orhan Firat, Dehao Chen, Mia~Xu
  Chen, HyoukJoong Lee, Jiquan Ngiam, Quoc~V Le, Yonghui Wu, et~al.
\newblock Gpipe: Efficient training of giant neural networks using pipeline
  parallelism.
\newblock In {\em NeurIPS}, 2019.

\bibitem[\protect\citeauthoryear{Isyanto \bgroup \em et al.\egroup
  }{2020}]{smart_speaks}
Haris Isyanto, Ajib~Setyo Arifin, and Muhammad Suryanegara.
\newblock Design and implementation of iot-based smart home voice commands for
  disabled people using google assistant.
\newblock In {\em 2020 International Conference on Smart Technology and
  Applications (ICoSTA)}, pages 1--6. IEEE, 2020.

\bibitem[\protect\citeauthoryear{Kang \bgroup \em et al.\egroup
  }{2017}]{Neurosurgeon}
Yiping Kang, Johann Hauswald, Cao Gao, Austin Rovinski, Trevor Mudge, Jason
  Mars, and Lingjia Tang.
\newblock Neurosurgeon: Collaborative intelligence between the cloud and mobile
  edge.
\newblock {\em SIGARCH Comput. Archit. News}, 45(1):615–629, April 2017.

\bibitem[\protect\citeauthoryear{Kim \bgroup \em et al.\egroup
  }{2020}]{torchgpipe}
Chiheon Kim, Heungsub Lee, Myungryong Jeong, Woonhyuk Baek, Boogeon Yoon, Ildoo
  Kim, Sungbin Lim, and Sungwoong Kim.
\newblock torchgpipe: On-the-fly pipeline parallelism for training giant
  models.
\newblock {\em CoRR}, abs/2004.09910, 2020.

\bibitem[\protect\citeauthoryear{Krizhevsky and
  others}{2012}]{krizhevsky2012imagenet}
Alex Krizhevsky et~al.
\newblock {ImageNet} classification with deep convolutional neural networks.
\newblock In {\em Advances in Neural Information Processing Systems}, pages
  1097--1105, 2012.

\bibitem[\protect\citeauthoryear{Kundu and
  Sundaresan}{2021}]{kundu2021attentionlite}
Souvik Kundu and Sairam Sundaresan.
\newblock Attentionlite: Towards efficient self-attention models for vision.
\newblock In {\em ICASSP 2021-2021 IEEE International Conference on Acoustics,
  Speech and Signal Processing (ICASSP)}, pages 2225--2229. IEEE, 2021.

\bibitem[\protect\citeauthoryear{Kundu \bgroup \em et al.\egroup
  }{2020}]{kundu2020pre}
Souvik Kundu, Mahdi Nazemi, Massoud Pedram, Keith~M Chugg, and Peter~A Beerel.
\newblock Pre-defined sparsity for low-complexity convolutional neural
  networks.
\newblock {\em IEEE Transactions on Computers}, 69(7):1045--1058, 2020.

\bibitem[\protect\citeauthoryear{Kundu \bgroup \em et al.\egroup
  }{2021}]{kundu2021dnr}
Souvik Kundu, Mahdi Nazemi, Peter~A Beerel, and Massoud Pedram.
\newblock Dnr: A tunable robust pruning framework through dynamic network
  rewiring of dnns.
\newblock In {\em Proceedings of the 26th Asia and South Pacific Design
  Automation Conference}, pages 344--350, 2021.

\bibitem[\protect\citeauthoryear{Li \bgroup \em et al.\egroup
  }{2021}]{li2021terapipe}
Zhuohan Li, Siyuan Zhuang, Shiyuan Guo, Danyang Zhuo, Hao Zhang, Dawn Song, and
  Ion Stoica.
\newblock Terapipe: Token-level pipeline parallelism for training large-scale
  language models.
\newblock {\em arXiv preprint arXiv:2102.07988}, 2021.

\bibitem[\protect\citeauthoryear{Liu \bgroup \em et al.\egroup }{2021}]{vec}
Lei Liu, Chen Chen, Qingqi Pei, Sabita Maharjan, and Yan Zhang.
\newblock Vehicular edge computing and networking: A survey.
\newblock {\em Mobile Networks and Applications}, 26(3):1145--1168, 2021.

\bibitem[\protect\citeauthoryear{Mao \bgroup \em et al.\egroup
  }{2017}]{mao2017modnn}
Jiachen Mao, Xiang Chen, Kent~W Nixon, Christopher Krieger, and Yiran Chen.
\newblock Modnn: Local distributed mobile computing system for deep neural
  network.
\newblock In {\em Design, Automation \& Test in Europe Conference \& Exhibition
  (DATE), 2017}, pages 1396--1401. IEEE, 2017.

\bibitem[\protect\citeauthoryear{Narayanan \bgroup \em et al.\egroup
  }{2019}]{pipedream}
Deepak Narayanan, Aaron Harlap, Amar Phanishayee, Vivek Seshadri, Nikhil
  Devanur, Greg Granger, Phil Gibbons, and Matei Zaharia.
\newblock Pipedream: Generalized pipeline parallelism for dnn training.
\newblock In {\em ACM Symposium on Operating Systems Principles (SOSP 2019)},
  October 2019.

\bibitem[\protect\citeauthoryear{Narayanan \bgroup \em et al.\egroup
  }{2021a}]{pipedream_2bw}
Deepak Narayanan, Amar Phanishayee, Kaiyu Shi, Xie Chen, and Matei Zaharia.
\newblock Memory-efficient pipeline-parallel dnn training.
\newblock In {\em 2021 International Conference on Machine Learning (ICML
  2021)}, July 2021.

\bibitem[\protect\citeauthoryear{Narayanan \bgroup \em et al.\egroup
  }{2021b}]{narayanan2021efficient}
Deepak Narayanan, Mohammad Shoeybi, Jared Casper, Patrick LeGresley, Mostofa
  Patwary, Vijay~Anand Korthikanti, Dmitri Vainbrand, Prethvi Kashinkunti,
  Julie Bernauer, Bryan Catanzaro, Amar Phanishayee, and Matei Zaharia.
\newblock Efficient large-scale language model training on gpu clusters, 2021.

\bibitem[\protect\citeauthoryear{Park \bgroup \em et al.\egroup
  }{2020a}]{hetpipe}
Jay~H Park, Gyeongchan Yun, M~Yi Chang, Nguyen~T Nguyen, Seungmin Lee, Jaesik
  Choi, Sam~H Noh, and Young-ri Choi.
\newblock Hetpipe: Enabling large $\{$DNN$\}$ training on (whimpy)
  heterogeneous $\{$GPU$\}$ clusters through integration of pipelined model
  parallelism and data parallelism.
\newblock In {\em 2020 $\{$USENIX$\}$ Annual Technical Conference
  ($\{$USENIX$\}$$\{$ATC$\}$ 20)}, pages 307--321, 2020.

\bibitem[\protect\citeauthoryear{Park \bgroup \em et al.\egroup
  }{2020b}]{park2020optimus}
Junki Park, Hyunsung Yoon, Daehyun Ahn, Jungwook Choi, and Jae-Joon Kim.
\newblock Optimus: Optimized matrix multiplication structure for transformer
  neural network accelerator.
\newblock {\em Proceedings of Machine Learning and Systems}, 2:363--378, 2020.

\bibitem[\protect\citeauthoryear{Paszke \bgroup \em et al.\egroup
  }{2019}]{NEURIPS2019_9015}
Adam Paszke, Sam Gross, Francisco Massa, Adam Lerer, James Bradbury, Gregory
  Chanan, Trevor Killeen, Zeming Lin, Natalia Gimelshein, Luca Antiga, Alban
  Desmaison, Andreas Kopf, Edward Yang, Zachary DeVito, Martin Raison, Alykhan
  Tejani, Sasank Chilamkurthy, Benoit Steiner, Lu~Fang, Junjie Bai, and Soumith
  Chintala.
\newblock Pytorch: An imperative style, high-performance deep learning library.
\newblock In {\em Advances in Neural Information Processing Systems 32}, pages
  8024--8035. 2019.

\bibitem[\protect\citeauthoryear{Premsankar \bgroup \em et al.\egroup
  }{2018}]{latency}
Gopika Premsankar, Mario Di~Francesco, and Tarik Taleb.
\newblock Edge computing for the internet of things: A case study.
\newblock {\em IEEE Internet of Things Journal}, 5(2):1275--1284, 2018.

\bibitem[\protect\citeauthoryear{Redmon and Farhadi}{2017}]{redmon2017yolo9000}
Joseph Redmon and Ali Farhadi.
\newblock Yolo9000: better, faster, stronger.
\newblock In {\em Proceedings of the IEEE conference on computer vision and
  pattern recognition}, pages 7263--7271, 2017.

\bibitem[\protect\citeauthoryear{Satyanarayanan}{2017}]{edgecomputing}
Mahadev Satyanarayanan.
\newblock The emergence of edge computing.
\newblock {\em Computer}, 50(1):30--39, 2017.

\bibitem[\protect\citeauthoryear{Sharma \bgroup \em et al.\egroup
  }{2021}]{wireless_sensor_network}
Himanshu Sharma, Ahteshamul Haque, and Frede Blaabjerg.
\newblock Machine learning in wireless sensor networks for smart cities: A
  survey.
\newblock {\em Electronics}, 10(9), 2021.

\bibitem[\protect\citeauthoryear{Shoeybi \bgroup \em et al.\egroup
  }{2019}]{Megatron-LM2019}
Mohammad Shoeybi, Mostofa Patwary, Raul Puri, Patrick LeGresley, Jared Casper,
  and Bryan Catanzaro.
\newblock Megatron-lm: Training multi-billion parameter language models using
  model parallelism.
\newblock {\em CoRR}, abs/1909.08053, 2019.

\bibitem[\protect\citeauthoryear{Su \bgroup \em et al.\egroup
  }{2019}]{su2019pixel}
Hang Su, Varun Jampani, Deqing Sun, Orazio Gallo, Erik Learned-Miller, and Jan
  Kautz.
\newblock Pixel-adaptive convolutional neural networks.
\newblock In {\em Proceedings of the IEEE/CVF Conference on Computer Vision and
  Pattern Recognition}, pages 11166--11175, 2019.

\bibitem[\protect\citeauthoryear{Tambe \bgroup \em et al.\egroup
  }{2020}]{tambe2020edgebert}
Thierry Tambe, Coleman Hooper, Lillian Pentecost, Tianyu Jia, En-Yu Yang, Marco
  Donato, Victor Sanh, Paul Whatmough, Alexander~M Rush, David Brooks, et~al.
\newblock Edgebert: Sentence-level energy optimizations for latency-aware
  multi-task nlp inference.
\newblock {\em arXiv preprint arXiv:2011.14203}, 2020.

\bibitem[\protect\citeauthoryear{Tao \bgroup \em et al.\egroup
  }{2018}]{tao2018image}
Hu~Tao, Weihua Li, Xianxiang Qin, and Dan Jia.
\newblock Image semantic segmentation based on convolutional neural network and
  conditional random field.
\newblock In {\em 2018 Tenth International Conference on Advanced Computational
  Intelligence (ICACI)}, pages 568--572. IEEE, 2018.

\bibitem[\protect\citeauthoryear{Touvron \bgroup \em et al.\egroup
  }{2021}]{deit2021}
Hugo Touvron, Matthieu Cord, Matthijs Douze, Francisco Massa, Alexandre
  Sablayrolles, and Herv{\'e} J{\'e}gou.
\newblock Training data-efficient image transformers \& distillation through
  attention.
\newblock In {\em International Conference on Machine Learning}, pages
  10347--10357. PMLR, 2021.

\bibitem[\protect\citeauthoryear{Vaswani \bgroup \em et al.\egroup
  }{2017}]{vaswani2017attention}
Ashish Vaswani, Noam Shazeer, Niki Parmar, Jakob Uszkoreit, Llion Jones,
  Aidan~N Gomez, Lukasz Kaiser, and Illia Polosukhin.
\newblock Attention is all you need.
\newblock {\em arXiv preprint arXiv:1706.03762}, 2017.

\bibitem[\protect\citeauthoryear{Wang \bgroup \em et al.\egroup }{2020}]{hat}
Hanrui Wang, Zhanghao Wu, Zhijian Liu, Han Cai, Ligeng Zhu, Chuang Gan, and
  Song Han.
\newblock Hat: Hardware-aware transformers for efficient natural language
  processing.
\newblock In {\em Annual Conference of the Association for Computational
  Linguistics}, 2020.

\bibitem[\protect\citeauthoryear{Wolf \bgroup \em et al.\egroup
  }{2020}]{wolf-etal-2020-transformers}
Thomas Wolf, Lysandre Debut, Victor Sanh, Julien Chaumond, Clement Delangue,
  Anthony Moi, Pierric Cistac, Tim Rault, Rémi Louf, Morgan Funtowicz, Joe
  Davison, Sam Shleifer, Patrick von Platen, Clara Ma, Yacine Jernite, Julien
  Plu, Canwen Xu, Teven~Le Scao, Sylvain Gugger, Mariama Drame, Quentin Lhoest,
  and Alexander~M. Rush.
\newblock Transformers: State-of-the-art natural language processing.
\newblock In {\em Proceedings of the 2020 Conference on Empirical Methods in
  Natural Language Processing: System Demonstrations}, pages 38--45, Online,
  October 2020.

\bibitem[\protect\citeauthoryear{Wu \bgroup \em et al.\egroup
  }{2019}]{wu2019lite}
Zhanghao Wu, Zhijian Liu, Ji~Lin, Yujun Lin, and Song Han.
\newblock Lite transformer with long-short range attention.
\newblock In {\em International Conference on Learning Representations}, 2019.

\bibitem[\protect\citeauthoryear{Yang \bgroup \em et al.\egroup
  }{2021}]{pipemare}
Bowen Yang, Jian Zhang, Jonathan Li, Christopher Re, Christopher Aberger, and
  Christopher De~Sa.
\newblock Pipemare: Asynchronous pipeline parallel dnn training.
\newblock In A.~Smola, A.~Dimakis, and I.~Stoica, editors, {\em Proceedings of
  Machine Learning and Systems}, volume~3, pages 269--296, 2021.

\bibitem[\protect\citeauthoryear{Yao \bgroup \em et al.\egroup
  }{2021}]{yao2021hawq}
Zhewei Yao, Zhen Dong, Zhangcheng Zheng, Amir Gholami, Jiali Yu, Eric Tan,
  Leyuan Wang, Qijing Huang, Yida Wang, Michael Mahoney, et~al.
\newblock Hawq-v3: Dyadic neural network quantization.
\newblock In {\em International Conference on Machine Learning}, pages
  11875--11886. PMLR, 2021.

\bibitem[\protect\citeauthoryear{Yuan \bgroup \em et al.\egroup
  }{2021}]{yuan2021tokens}
Li~Yuan, Yunpeng Chen, Tao Wang, Weihao Yu, Yujun Shi, Zihang Jiang, Francis~EH
  Tay, Jiashi Feng, and Shuicheng Yan.
\newblock Tokens-to-token vit: Training vision transformers from scratch on
  imagenet.
\newblock {\em arXiv preprint arXiv:2101.11986}, 2021.

\bibitem[\protect\citeauthoryear{Zhang \bgroup \em et al.\egroup
  }{2018}]{zhang2018systematic}
Tianyun Zhang, Shaokai Ye, Kaiqi Zhang, Jian Tang, Wujie Wen, Makan Fardad, and
  Yanzhi Wang.
\newblock A systematic dnn weight pruning framework using alternating direction
  method of multipliers.
\newblock In {\em Proceedings of the European Conference on Computer Vision
  (ECCV)}, pages 184--199, 2018.

\bibitem[\protect\citeauthoryear{Zhao \bgroup \em et al.\egroup
  }{2018}]{zhao2018deepthings}
Zhuoran Zhao, Kamyar~Mirzazad Barijough, and Andreas Gerstlauer.
\newblock Deepthings: Distributed adaptive deep learning inference on
  resource-constrained iot edge clusters.
\newblock {\em IEEE Transactions on Computer-Aided Design of Integrated
  Circuits and Systems}, 37(11):2348--2359, 2018.

\bibitem[\protect\citeauthoryear{Zhou \bgroup \em et al.\egroup
  }{2019}]{AdaptiveExecution}
Li~Zhou, Mohammad~Hossein Samavatian, Anys Bacha, Saikat Majumdar, and Radu
  Teodorescu.
\newblock Adaptive parallel execution of deep neural networks on heterogeneous
  edge devices.
\newblock New York, NY, USA, 2019. Association for Computing Machinery.

\end{thebibliography}
\end{document}